\documentclass{aastex62}

\usepackage{amsmath,amssymb,amsfonts}
\usepackage{algorithmic}
\usepackage{textcomp}
\usepackage{booktabs}

\newcommand*{\vertbar}{\rule[-1ex]{0.5pt}{2.5ex}}

\newcommand{\Cov}{\mathrm{Cov}}

\DeclareMathOperator*{\argmax}{arg\,max}
\newcommand{\LRT}[2]{%
    \mathrel{\mathop\gtrless\limits^{#1}_{#2}}%
}

\graphicspath{{./}{figures/}}

\received{October 6, 2019}
\accepted{\today}
\submitjournal{ApJ}

\shorttitle{Bayesian Joint Detection of Transiting Exoplanets}
\shortauthors{Taaki, Kamalabadi, \& Kemball}

\begin{document}

\title{Bayesian Methods for Joint Exoplanet Transit Detection and Systematic Noise Characterization}

\correspondingauthor{Jamila Taaki}
\email{xiaziyna@gmail.com}

\author[0000-0001-5475-1975]{Jamila S. Taaki}
\affiliation{Department of Electrical and Computer Engineering,
University of Illinois at Urbana-Champaign\\
306 N. Wright St. MC 702, Urbana, IL 61801-2918}

\author{Farzad Kamalabadi}
\affiliation{Department of Electrical and Computer Engineering,
University of Illinois at Urbana-Champaign\\
306 N. Wright St. MC 702, Urbana, IL 61801-2918}

\author[0000-0001-6233-8347]{Athol J. Kemball}
\affiliation{Department of Astronomy, University of Illinois at Urbana-Champaign\\
1002 W. Green Street, Urbana, IL 61801-3074}
\nocollaboration



\begin{abstract}
The treatment of systematic noise is a significant aspect of transit exoplanet data processing due to the signal strength of systematic noise relative to a transit signal. Typically the standard approach to transit detection is to estimate and remove systematic noise independently of and prior to a transit detection test. If a transit signal is present in a light curve, the process of systematic noise removal may distort the transit signal by overfitting and thereby reduce detection efficiency. We present a Bayesian framework for joint detection of transit signals and systematic noise characterization and describe the implementation of these detectors as optimal Neyman-Pearson likelihood ratio tests. The joint detectors reduce to closed form as matched filters under the assumption of a Gaussian Bayesian prior for the systematic noise. The performance of the exploratory detectors was evaluated in injection tests and show $\sim 2\%$ improvement in overall detection efficiency relative to the standard approach. We find that joint detection efficiency is specifically improved for short-period, low transit-depth exoplanet transits, providing evidence in support of the hypothesis that joint detection may indeed help to mitigate overfitting. In addition, an initial feasibility test to detect known exoplanets in Kepler data using the joint detectors produced encouraging preliminary results.
\end{abstract}



\section{Introduction} \label{sec:intro}
Transiting exoplanet detection telescopes and missions such as CoRoT\footnote{sci.esa.int/corot} \citep{leger}, Kepler\footnote{keplerscience.arc.nasa.gov} \citep{Borucki977},
K2 \citep{Howell_2014} , and TESS\footnote{tess.mit.edu} \citep{ricker2014transiting} have been crucial to expanding the catalog of known exoplanets and their populations statistics. To date Kepler and K2 have produced approximately $\sim 3000$ confirmed exoplanet detections from $\sim 200,000$ observed light curves \citep{kepconf}. The detectability of a transiting exoplanet is limited by both the system performance of the telescope as well as the statistical efficiency of the detection and estimation methods used during data-processing. Continuous improvements in transit detection methods may reveal more exoplanets in existing datasets and also push the limits of observable exoplanet populations in current and future observations. In this paper we consider one such exploratory innovation in transit detection and estimation, as described in further detail below.
\par A transit signal is embedded in a variety of astrophysical \citep{fp} and instrumental signals \citep{noise}. Typical astrophysical noise signals tend to have known physical origin, such as photon counting noise, intrinsic stellar emission variability or variability from eclipsing systems, hence such noise contributions are often well described by deterministic \citep{torres2010modeling} or stochastic models \citep{scargle,borucki}. Noise arising from systematic error (hereinafter systematic noise) is instrument noise for which there exists no a priori data model and which is not reduced by averaging; it is often modeled non-parametrically. Some possible sources of systematic noise include unmodeled residual pointing errors \citep{foreman}, instrumental detector offsets, and seasonal instrumental variations. Systematic noise may have trends over a range of time-scales including short-duration or transient outlier events.
\par A large number of algorithms exist for the inference and removal of, systematic noise from wide-field transit telescope light curves. A detailed overview of a number of techniques is provided by \citet{roberts} who also provide their own algorithm for systematics removal. The prevalent approach is cotrending: a set of basis signals representative of a set of light curves is obtained and linear combinations of these basis signals are used to form net estimates of systematic noise; these are then removed from all light curves processed \citep{Kinemuchi_2012}. In the Trend-Filtering-Algorithm (TFA) \citep{tfa} the basis is chosen as a random subset of the light curves themselves. The method of principal component analysis (PCA) is an alternative method to create a set of basis signals from a large set of light curves; a basis obtained from PCA is orthogonal and maximizes the variance of the light curves projected onto it \citep{jolliffe2011principal}. The PCA method is used by the Sys-Rem detrending algorithm \citep{tamuz} and the Simultaneous Additive and Relative Systems Algorithm (SARS) \citep{ofir2010sars}. A similar related method, singular value decomposition (SVD) is used by the Kepler Pre-Search Data Conditioning (PDC) module \citep{twicken2010presearch, Stumpe_2012, kep}. The PDC-MAP (Maximum A-Posteriori) algorithm \citep{Stumpe_2012, kep} was later developed from PDC-LS (least-squares) \citep{twicken2010presearch}; the latter forms least-square systematic noise estimates from a set of basis signals. The PDC-MAP algorithm forms an empirical prior over the set of basis signals and forms Bayesian maximum a posteriori (MAP) systematic noise estimates.
\par Systematic noise estimates are typically formed without any assumption of whether a transit signal is or is not present in the underlying light curves. However the signal space of transits and systematic noise may not be unambiguously separated in either the time or frequency domain. For example, \citet{noise} discuss the temporal properties of early Kepler data and identify several short-timescale sources of systematic error in the photometric lightcurves. High-frequency ($\leq 10$ d) systematic noise is similarly identified in Kepler photometry by \citet{petigura2012identification}. This presents a challenge to sequential estimation of systematic noise and transit detection. If a transit signal is present in a lightcurve, the estimated systematic noise may be biased and the subsequent transit detection process may therefore be also adversely affected \citep{Christiansen_2013, foreman}. This point motivates joint modelling of systematics and transits.
\par In this paper we consider an exploratory Bayesian approach of jointly estimating systematic noise and transit signals with the goal of improving detection rates while reducing the false-alarm rate. This idea has been explored in a non-Bayesian setting; \citet{foreman} provide a method that finds joint linear maximum likelihood estimates of transit signals and systematic noise. The methods that we propose here are Bayesian in nature. The advantage of PDC-MAP over PDC-LS, as described above, has set a precedent for Bayesian systematic noise treatment that leads us to believe it is also advantageous in this setting. Frequentist estimation of systematic noise considers every systematic basis signal as a-priori equally likely of occurring in a light curve. Arguably this model is not a realistic description of an obtained basis and may produce poor systematic estimates. For example a basis obtained via PCA may be ordered by its singular values, providing a measure of the dominance of various basis signals within the set of light curves. Furthermore there is expected variability in the presence or relative influence of individual basis signals across a set of separate lightcurves. If this information is discarded, a basis signal which occurred in a small number, or in a particular subset, of light curves may be unrealistically included in estimates of other light curves, in turn this biased systematics estimate may lead to poor transit detection performance. In a Bayesian treatment such as PDC-MAP any ancillary information of this nature can be utilized to fully inform more realistic systematic noise estimates.
\par The work presented here on joint transit detection with a Bayesian treatment of systematic noise is summarized as follows. Two methods are derived to compute detection tests on raw light curves: i) the first method marginalizes over a Bayesian prior describing the systematic noise; while, ii)  the second method forms fixed Bayesian estimates of the systematic noise. The use of marginalization in (i) allows an averaged detection test to be computed over a continuous set of systematic noise models. Our second approach (ii) forms fixed Bayesian estimates of the systematic noise under two models in which the signal does or does not contain a transit signal; these are then used as input into a detection test. We describe a detection framework based on these two methods and derive analytic detectors. In the work by \citet{Luger_2017}, a Bayesian joint model of similar form is used to derive a computationally tractable likelihood function which the authors propose may be used for a variety of astrophysical purposes including transit light curve modelling. Their derivation follows a marginalization approach whereby the properties of Gaussianity lead to an analytic form for a likelihood function. In deriving Bayesian detection strategies we also examine a marginal Gaussian likelihood function which leads to a simple analytic detector. However the framework we propose and the derived detectors are general; further we formulate the detector as a binary hypothesis test.

\par PDC-MAP as included in version 8.0 of the Kepler science pipeline formed part of a continuous improvement in presearch data conditioning; those residual systematic errors remaining were believed to arise from the fact that the systematic basis vectors were not a statistically independent set \citep{Stumpe_2014}. This was remedied by introducing a refined multiscale PDC-MAP algorithm (msMAP) \citep{Stumpe_2014}. This method uses wavelet filtering to separate the different temporal time scales of the systematic effects; PDC-MAP is then used independently in each sub-band and the results synthesized to correct the lightcurve. Our algorithmic approach includes the ability to model correlations between systematic noise trends. We assume statistical correlations exist between systematic basis vectors and have a model prior that may include such information. This represents a complementary approach to that adopted by msMAP.

\par We demonstrate the performance of our detection methods by performing single-transit injection tests on Kepler data which exclude prior known exoplanet detections. We simulate a a set of limb-darkened transit signals enumerated in Table \ref{tab:transitparams}. In the injection tests the new joint Bayesian detection methods outperformed a standard sequential cotrending and detection approach with a relative average $\sim 3 \%$ improvement in the overall detection efficiency subject to the same rate of incorrect detections. In particular the injection tests demonstrated the ability of these techniques to separate short period transits ($<$ 10 days) in the presence of high-frequency systematics.  We also demonstrated the feasibility of these methods on a small subset of Kepler light curves when the sample included prior exoplanet detections. 
\par The paper is organized as follows. Section \ref{sec:detmodel} introduces the light curve signal model, the joint detection framework, particular detector implementations, and describes a set of numerical tests to evaluate detector performance. In Section \ref{sec:results} we report the results from our injection and feasibility tests with Kepler data. The results are discussed in Section \ref{sec:discussion} and conclusions presented in Section \ref{sec:conclusion}. 
\begin{figure}[ht!]
\plotone{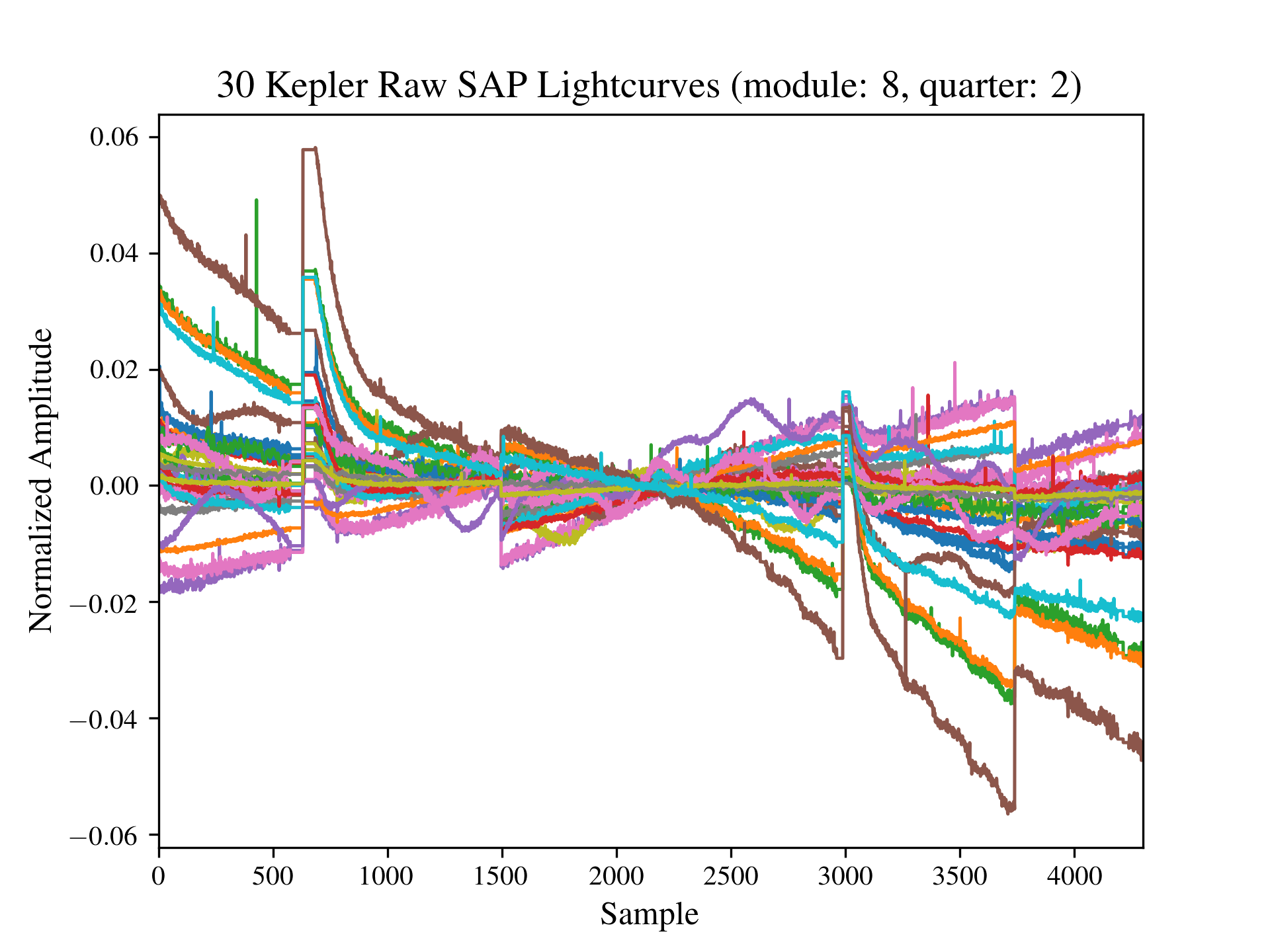}
\caption{A subset of Kepler raw simple aperture photometry (SAP) \citep{jenkins2010overview} light curves drawn from CCD module 8 and observing quarter 2, normalized and median subtracted. The light curves exhibit common trends due to systematic noise.}
\label{fig:samplelcs}
\end{figure}

\section{METHODS} \label{sec:detmodel}
Transit detection infers the presence of an exoplanet in orbit around a star from time series observations of an optically unresolved star-planet system; contemporary reviews are provided by \citet{deeg_alonso_2018} and \citet{transit}. As a planet obscures the face of a star, the observed brightness of the system drops. The effect a transiting planet has on the overall observed brightness of a star-planet system is well-approximated by parametric models \citep{Seager_2003}, Mandel-Agol equations \citep{mandel2002analytic}, or periodic box functions \citep{bls}, where the models and associated parameter sets depend largely on the properties of the star and planet. Expressed here in the language of detection theory, the transit method seeks to detect the presence of a transit signal of pre-specified functional form, but with unknown parameter values, in the observed light curve of a star-planet system.
\par Detection of a transit signal can be posed as a binary hypothesis test upon a candidate light curve: does the light curve contain a transit signal? Commonly this test is performed using a matched filter based test once cotrending
has been used to remove systematic noise from a light curve \citep{tingley2003rigorous}. Since we wish to jointly detect transit signals and model systematic noise in the current work we form the hypothesis test on raw light curves. A raw light curve $\mathbf{y}_i$ for a particular star indexed by $i \in I$ (where $I$ is the set of integers) is expressed as a vector of $N$ photometric flux measurements in time, before any cotrending for systematic noise, and hereinafter considered also to be normalized and median subtracted. An example subset of such raw light curves, here from the Kepler mission, is shown in Figure~\ref{fig:samplelcs}. 
The raw light curve vector $\mathbf{y}_i$ contains systematic noise $\mathbf{l}_i$ and a stellar signal $\mathbf{s}_i$. For conciseness, residual sources of non-systematic statistical error are considered included in $\mathbf{s}_i$. This includes instrumental shot noise, considered here in the Gaussian limit \citep{grinstead2012introduction}.

The test is as follows: the null hypothesis $(H_0)$ posits that $\mathbf{y}_i$ contains no transit signal. The alternative hypothesis $(H_1)$ posits that $\mathbf{y}_i$ does contain a transit signal $\mathbf{t} \in \mathbf{T}$. The set $\mathbf{T}$ describes all detectable transit signals. The hypothesis test is therefore posed as:
\begin{align}
H_0: \mathbf{y}_i &= \mathbf{s}_i + \mathbf{l}_i \\
H_1: \mathbf{y}_i &= \mathbf{t} + \mathbf{s}_i +  \mathbf{l}_i 
\label{eq:hypothesis}
\end{align}
To decide between hypotheses a likelihood ratio test \citep[LRT,][]{kaybook, wasserman_2013} compares the probability of the raw light curve $\mathbf{y}_i$ under either hypothesis model. If the LRT $\mathcal{L}_{i}(\mathbf{y}_i)$ exceeds a fixed threshold $\tau$ then a transit signal has been detected. More generally we consider the LRT as a test statistic $T({\mathbf{y}_i})$ on the data $\mathbf{y}_i$. Typically a Neyman-Pearson criterion \citep{kaybook, wasserman_2013}, in which the probability of detection is maximized subject to a fixed rate of false alarm $\alpha$, is used to decide the threshold $\tau$. These quantities are related as:
\begin{equation}
T({\mathbf{y}_i}) = \mathcal{L}_{i}(\mathbf{y}_i) = \frac{p_1(\mathbf{y}_i)}{p_0(\mathbf{y}_i) } \LRT{H_1}{H_0} \tau
\label{eq:lrt}
\end{equation}
where  $p_h(\mathbf{y}_i)  \equiv p(\mathbf{y}_i | H_h) $ is the probability of observing light curve $\mathbf{y}_i$ under a particular hypothesis model $h \in \{0,1 \}$.
\subsection{Raw Light Curve Signal Model}
This section describes typical models for the raw light curve components: the systematic noise, the stellar signal and the transit signal. 
\subsubsection{Systematic Noise}\label{sec:sysnoise}
A standard model for systematic noise is a linear reduced basis model \citep{tfa} in which systematic noise vectors $\mathbf{l}_i$ are modelled as a linear combination of a set of $K$ systematic noise basis vectors $\{ \mathbf{v}_k\}$ weighted by coefficients $c_i^k$. Since the set of systematic noise signals $\{ \mathbf{l}_i\}$ are reasonably expected to be correlated between light curves of sufficiently common instrumental origin \citep{Stumpe_2012}, they may be expressed in terms of a common set of systematic noise basis vectors $\{ \mathbf{v}_k\}$. 

\begin{equation}
\mathbf{l}_i = {\sum_{k=1}^{K} c_i^k \mathbf{v}_k}
\label{eq:li}
\end{equation}
Systematic basis noise vectors $\{ \mathbf{v}_k\}$ may be estimated via dimensionality reduction techniques \citep{cunningham2015linear} applied on $\{ \mathbf{y}_i \}$. Principal component analysis (PCA) is a commonly-used technique for cotrending and produces a set of orthogonal basis vectors \citep{sysrem, twicken2010presearch, Stumpe_2012, kep, petigura2012identification, foreman}.
A non-Bayesian least-squares estimation of the coefficients $\mathbf{c}_i$ is equivalent to maximum likelihood estimation \citep{kaymodern} with an assumed Gaussian stellar noise model $\mathbf{s}_i$ where the samples are independent and identically distributed. However, least-squares estimation, by minimizing the net root mean square (rms) error, is prone to overfitting residual transit signatures \citep{kep,Stumpe_2012}. As shown by \citet{kep} a Bayesian estimation of $\mathbf{c}_i$ can mitigate this issue by allowing incorporation of a prior $ p(\mathbf{c}_i)$ on the coefficients. The prior here is conditioned on the index $i \in I$; this index accommodates latent variables such as position within the CCD and the stellar magnitude, that produce clustering in systematic noise properties as noted above \citep{kep,Stumpe_2012}.

By definition, the cotrending model for systematic noise in Equation \ref{eq:li} does not fully capture systematic noise that is temporally uncorrelated between light curves. Examples of such systematic noise effects include cosmic ray events, sudden pixel sensitivity dropoff, and electronic image artifacts \citep{noise,Stumpe_2012}. Whilst our model cannot account for outlier effects, a number of residual unmodeled signatures may be treated as effectively Gaussian and absorbed into the term $\mathbf{s}_i$ in Equation \ref{eq:hypothesis}.
\subsubsection{Stellar Signal}\label{sec:stellar}
In the current work we assume that the stellar noise $\mathbf{s}_i$ (in each lightcurve $i$) may be reasonably modelled by a Gaussian noise model $\mathbf{s}_i \sim \mathcal{N}(\mathbf{0}, \Cov_{s,i})$. Gaussian processes are reviewed in the monograph by \citet[Ch.3]{gallager2013stochastic}. Gaussian models are powerful non-parametric models of processes with complex or unknown generating phenomena, and may be justified by the central limit theorem \citep{grinstead2012introduction}. Gaussian models have been shown to be effective for modelling stellar variability \citep{stelnoise} and have been used in the context of exoplanet detection \citep{carter2009parameter, rajpaul2015gaussian}.

Stellar flux density time-series display variability on a wide range of timescales \citep{conroy2018complete} and time-correlated fluctuations significantly affect the detectability of transit signals \citep{borucki, jenkins, pont}. A Gaussian noise model can incorporate time-correlated structure via its covariance matrix. A review of common correlated noise estimators for exoplanet lightcurves is provided by \citet{cubillos2016correlated}.

Our model further assumes that stellar noise is stationary within a fixed time window. In a stationary Gaussian colored noise model\footnote{A stationary Gaussian noise model is equivalent to a wide-sense-stationary Gaussian noise model, since Gaussian models are fully parameterized by their first and second moments \citep{papoulis2002}. }, correlations only depend on the separation between two points in time. Equivalently the values on the stellar covariance matrix $\Cov_{s, i}$ are constant along the diagonals and the matrix is Toeplitz in form. Such a model admits a power spectral representation and allows spectral analysis \cite{kaymodern}, which is considerably more efficient than time-domain analysis.

Stellar processes do however evolve in time and can exhibit non-stationarity as exemplified by our sun \citep{borucki, jenkins}. For this reason we confine our stellar noise model to be stationary only within a single observational quarter of Kepler data (approximately 90 days). This assumption is motivated in part by the analysis of common stellar periodic variability in Kepler data on time scales of days to weeks \citep{Basri_2010}.

The extension of our detectors to multiple Kepler observing quarters and associated implications for our statistical assumptions regarding signal covariance time scales are described in Appendix \ref{sec:multi-quarter}.

\subsubsection{Transit Signal} \label{sec:transit}
A transit signal is typically approximated as a periodic box function \citep{bls}. The periodic box function is parameterized as follows: $\alpha$ describes the fractional drop in light relative to the stellar signal, $P$ is the orbital period of the planet, $d$ is the transit duration, $n$ is the time index, and $t_0$ is the phase (epoch) of the signal. 
\begin{equation}
t_{\alpha, P, t_0, d} [n] = 
\begin{cases}
     \alpha ,& \text{if } (n-t_0)\mod_P \leq d \\
    0,              & \text{otherwise}
\end{cases} \label{eq:transit}
\end{equation}
A typical technique to estimate transit signal parameters is to quantize the space of possible parameter values to create a discrete set of candidate transit signals $\mathbf{T}$ and to search through this set while performing detection tests \citep{jenkins_2002}. The set of possible transit signals $\mathbf{T}$ is not described herein by a Bayesian prior as a-priori the population statistics of exoplanets are not fully known. As noted above other parametric forms can be used in place of a periodic box function \citep{Seager_2003, mandel2002analytic}.
\subsubsection{Signal Matrix Form}
The signal model may be expressed in matrix form. A similar form is presented in \citet{kep} which we adapt to include the presence of a transit signal.
As before time is indexed by $n$ ranging from $1$ to $N$ and $i$ is the light curve index. $\mathbf{V}$  is an $N\times K$ matrix with columns formed from $ \{ \mathbf{v}_k \}$:
\begin{equation} \mathbf{y}_i = \mathbf{t} +  \mathbf{V} \mathbf{c}_i + \mathbf{s}_i
\end{equation}
\begin{equation} \begin{bmatrix}
           {y}_{i}[1] \\
           {y}_{i}[2] \\
           \vdots \\
           {y}_{i}[N]
         \end{bmatrix}= 
         \begin{bmatrix}
           t[1] \\
           t[2] \\
           \vdots \\
           t[N]
         \end{bmatrix}
         + \begin{bmatrix}
     \vertbar & \vertbar &        & \vertbar \\
    \mathbf{v}_1    & \mathbf{v}_2    & \ldots & \mathbf{v}_K    \\
    \vertbar & \vertbar &        & \vertbar 
  \end{bmatrix} \begin{bmatrix}
           c_i^1 \\
           c_i^2 \\
           \vdots \\
           c_i^K
         \end{bmatrix} + \begin{bmatrix}
           s_i[1] \\
           s_i[2] \\
           \vdots \\
           s_i[N]
         \end{bmatrix}
\end{equation}
\subsection{Joint Detection Methods}\label{sec:jointdet}
In order to compute the LRT as described in Equation \ref{eq:lrt} it is necessary to compute $p_h(\mathbf{y}_i): h \in \{0, 1 \}$, the probability of a light curve conditioned on a hypothesis model. Under $H_1$ the light curve $\mathbf{y}_i$ contains a transit signal $\mathbf{t} \in \mathbf{T}$, where we assume that each possible transit signal is a-priori equally likely to occur. Hence for the remainder of this section we consider the LRT to be computed with respect to a fixed transit signal $\mathbf{t}$.
\par With the transit signal fixed, the signal model describing $\mathbf{y}_i$ for either hypothesis contains two unknown components: the systematic noise $\mathbf{l}_i = \mathbf{V}\mathbf{c}_i$ and the stellar signal $\mathbf{s}_i$. Computing the conditional likelihood of a light curve $\mathbf{y}_i$ for a particular $\mathbf{t}$, $p_h(\mathbf{y}_i | \mathbf{t}): h \in \{0, 1 \}$ requires addressing the dependence on these unknown terms. We use two standard approaches in Bayesian analysis for this purpose, namely marginalization over the systematics and estimation of the systematic noise. This results in two distinct detection methodologies.
\par With Bayesian marginalization \citep{loredo1992promise}, the joint probability distribution describing the observations and signal model is integrated with respect to the systematic noise to produce two marginal likelihoods for either hypothesis model. The ratio of these marginal likelihoods forms the LRT.
In the second approach, systematics are estimated conditioned on the hypothesis model. This amounts to finding two distinct Bayesian systematics estimates both under the belief that there is a transit signal in the light curve and that there is no transit signal. These fixed systematics estimates are used to compute the likelihoods $p_h(\mathbf{y}_i): h \in \{0, 1 \}$, the ratio of which forms the LRT.
\par This section derives generic forms of the LRT obtained through Bayesian marginalization and fixed estimation. Specific implementation details are provided in following sections.

\subsubsection{Matched Filter}\label{sec:matched_filter}
Before proceeding further, a brief overview of the matched filter is provided as it plays an essential part in our detection framework. For more detail we refer the reader to the monographs by \citet{kaybook} and \citet{poor2013introduction}. 
\par The matched filter is the optimal Neyman-Pearson detector for a deterministic signal $\mathbf{t}$ in Gaussian noise $\mathbf{n} \thicksim \mathcal{N} (\mathbf{0}, \Cov_n)$ for an observed signal $\mathbf{y}$. Using the form provided by \citet{jenkins_2002} the matched filter test statistic $T(\mathbf{y})$  is given by:
\begin{align}
T (\mathbf{y}) = \frac{\mathbf{y}^T \Cov_n^{-1} \mathbf{t}}{ \sqrt{\mathbf{t}^T\Cov_n^{-1} \mathbf{t}}} \LRT{H_1}{H_0} \tau
\label{eq:matchedfilter}
\end{align}
\par This form of the matched filter may be derived from the LRT formulation in Equation \ref{eq:lrt} under the assumption of Gaussianity. The matched filter is desirable from an implementation standpoint as the distribution of the matched filter test statistic conditioned on the null hypothesis is normal: $T(\mathbf{y})|H_0 \sim \mathcal{N}(0, 1)$. Thus a detection threshold $\tau$ should achieve the same false-alarm rate for different noise covariance matrices $\Cov_n$  and signals $\mathbf{t}$. Secondly if $\mathbf{n}$ is wide-sense-stationary, the matched filter may be efficiently computed in the Fourier domain \citep{kay}.
\subsubsection{Detector A: Marginalization over Systematic Noise}\label{sec:deta}
This detector computes $p_h(\mathbf{y}_i): h \in \{0, 1 \}$ by marginalizing over the systematic noise prior probability $p(\mathbf{c}_i)$. For a particular hypothesis, the conditional probability of the observed signal $\mathbf{y}_i$ conditioned on systematic noise coefficients $\mathbf{c}_i$ is given by:
\begin{align}
p_0(\mathbf{y}_i | \mathbf{c}_i ) = p( \mathbf{s}_i = \mathbf{y}_i - \mathbf{V} \mathbf{c}_i  | \mathbf{c}_i ) \label{eq:detb_pzero}\\
p_1(\mathbf{y}_i |\mathbf{c}_i) = p(\mathbf{s}_i = \mathbf{y}_i - \mathbf{t} - \mathbf{V} \mathbf{c}_i | {\mathbf{c}_i}) \label{eq:detb_pone}
\end{align}
\par Since the $\mathbf{t}$ and $\mathbf{y}_i$ are deterministic, the only stochastic term in these distributions is $\mathbf{s}_i$ which is modelled as approximately Gaussian. See Sections \ref{sec:detmodel} and \ref{sec:sysnoise} for a discussion of contributing non-Gaussian terms in $\mathbf{s}_i$. The LRT may be computed by marginalizing over the known prior $p(\mathbf{c}_i)$ for both of these conditional probability functions:
\begin{align}
\mathcal{L}_{i}(\mathbf{y}_i) = \frac{p_1(\mathbf{y}_i)}{p_0(\mathbf{y}_i) } =\frac{\int_{\mathbf{c}_i \in \mathbf{C}_i} p_1(\mathbf{y}_i | \mathbf{c}_i ) p(\mathbf{c}_i)\ d\mathbf{c}_i}{\int_{\mathbf{c}_i \in \mathbf{C}_i} p_0(\mathbf{y}_i | \mathbf{c}_i ) p(\mathbf{c}_i)\ d\mathbf{c}_i }
\\=\frac{\int_{\mathbf{c}_i \in \mathbf{C}_i} p( \mathbf{y}_i = \mathbf{t} + \mathbf{V} \mathbf{c}_i + \mathbf{s}_i | {\mathbf{c}_i}) p(\mathbf{c}_i)\ d\mathbf{c}_i}
{\int_{\mathbf{c}_i \in \mathbf{C}_i} p(\mathbf{y}_i =\mathbf{V} \mathbf{c}_i + \mathbf{s}_i| {\mathbf{c}_i}) p(\mathbf{c}_i)\ d\mathbf{c}_i} \label{eq:deta_lrt}
\end{align}
where $\mathbf{C}_i$ is the domain of $\mathbf{c}_i$. 

\subsubsection{Detector B: Joint Transit and Systematic Noise Estimation}\label{sec:detb}

This method obtains two fixed systematic noise estimates $\mathbf{\hat{c}}_1$ and $\mathbf{\hat{c}}_0$, with and without the presence of a transit signal respectively. These Bayesian estimates are obtained from the posterior distributions $p_{h}(\mathbf{c}_i|\mathbf{y}_i) : h \in \{0, 1 \}$ currently as maximum-a-posteriori (MAP) estimates. These can then be used to compute  $p_h(\mathbf{y}_i): h \in \{0, 1 \}$ and the LRT. Herein lies the main distinction from prior cotrending approaches; there a single systematic noise estimate, roughly equivalent to $\mathbf{\hat{c}}_0$, is used to compute a detection test. Cotrending is performed before detection and the systematic noise estimate is formed without modelling a particular transit signal. Such an approach favors the null hypothesis in a detection test as $\mathbf{\hat{c}}_0$ maximizes the null hypothesis posterior. Here instead we find the MAP estimates for systematic noise under both transit signal hypotheses and compare the likelihoods computed from these estimates in the form of the LRT.
\par The conditional MAP estimates of the systematic noise under either hypothesis take the form:
\begin{align}
\mathbf{\hat{c}^{MAP}_{1,i}}= \argmax_{\mathbf{c}_i \in \mathbf{C}_i} p_{1}(\mathbf{c}_i|\mathbf{y}_i) =
\argmax_{\mathbf{c}_i \in \mathbf{C}_i} p_{1}(\mathbf{y}_i|\mathbf{c}_i) p(\mathbf{c}_i) \label{eq:detb_c1map} \\
= \argmax_{\mathbf{c}_i \in \mathbf{C}_i} \left( \ln[p_1(\mathbf{y}_i|\mathbf{c}_i)] + \ln[p(\mathbf{c}_i)] \right)
\\
\mathbf{\hat{c}^{MAP}_{0,i}} = \argmax_{\mathbf{c}_i \in \mathbf{C}_i} \left( \ln[p_0(\mathbf{y}_i|\mathbf{c}_i)] + \ln[p(\mathbf{c}_i)] \right) 
\label{eq:detb_c0map}
\end{align}
\par We insert these estimates into the hypothesis test described in Equation \ref{eq:hypothesis} as:
\begin{align}
 H_0: \mathbf{y}_i &= \mathbf{V}\mathbf{\hat{c}^{MAP}_{0,i}} + \mathbf{s}_i \label{eq:detb_h0} \\
 H_1: \mathbf{y}_i &= \mathbf{t} + \mathbf{V}\mathbf{\hat{c}^{MAP}_{1,i}} + \mathbf{s}_i \end{align}
\par Since the systematic noise signal for either hypothesis is fixed the only stochastic variable is the Gaussian stellar noise $\mathbf{s}_i$. The detection test may be equivalently restated as:
\begin{align}
H_0: \mathbf{\hat{y_i}} &=  \mathbf{s}_i \\
H_1: \mathbf{\hat{y_i}} &= \mathbf{k}_i + \mathbf{s}_i 
\end{align}
\par This hypothesis test on $\mathbf{\hat{y_i}} = \mathbf{y}_i - \mathbf{V} \mathbf{\hat{c}^{MAP}_{0,i}}$ describes detection of a known signal $\mathbf{k}_i =  \mathbf{t} - \mathbf{V} \mathbf{\hat{c}^{MAP}_{0,i}} + \mathbf{V} \mathbf{\hat{c}^{MAP}_{1,i}}$ in Gaussian noise $\mathbf{s}_i$. Therefore, as described above in Section \ref{sec:matched_filter}, the optimal detector is the matched filter with test statistic $T_i (\mathbf{\hat{y_i}})$:
\begin{align}
T_i (\mathbf{\hat{y_i}}) = \frac{\mathbf{\hat{y}}_i^T \Cov_{s,i}^{-1} \mathbf{k}_i}{ \sqrt{\mathbf{k}_i^T\Cov_{s,i}^{-1} \mathbf{k}_i}} \LRT{H_1}{H_0} \tau
\label{eq:detb}
\end{align}
\subsection{Design and Implementation of the Joint Detectors}\label{sec:design_impl}
The detectors defined in Section \ref{sec:jointdet} admit several choices in their concrete implementation, including assumptions regarding the statistical distribution of the underlying variables defining the systematic noise and stellar signal and the inference methods used to estimate their parameter values. In this regard our design and implementation decisions have been wherever possible scientifically or empirically motivated and guided also by computational feasibility. The assumed statistical distribution of the systematic noise coefficient vector $\mathbf{c}_i$ as well as its dimension ($K$) largely determine the analytic and computational tractability of the detectors. A small number of distributions produce closed form posterior distributions and thereby analytic forms of these detectors; specifically the implementation for detectors A and B is described here under the assumption of Gaussian statistics. The inference methods used to estimate the detector parameter values are described at the close of this Section.

\subsubsection{Systematic Noise Prior: Gaussianity} \label{sec:gaussian_sys}
\begin{figure}[ht!]
\plotone{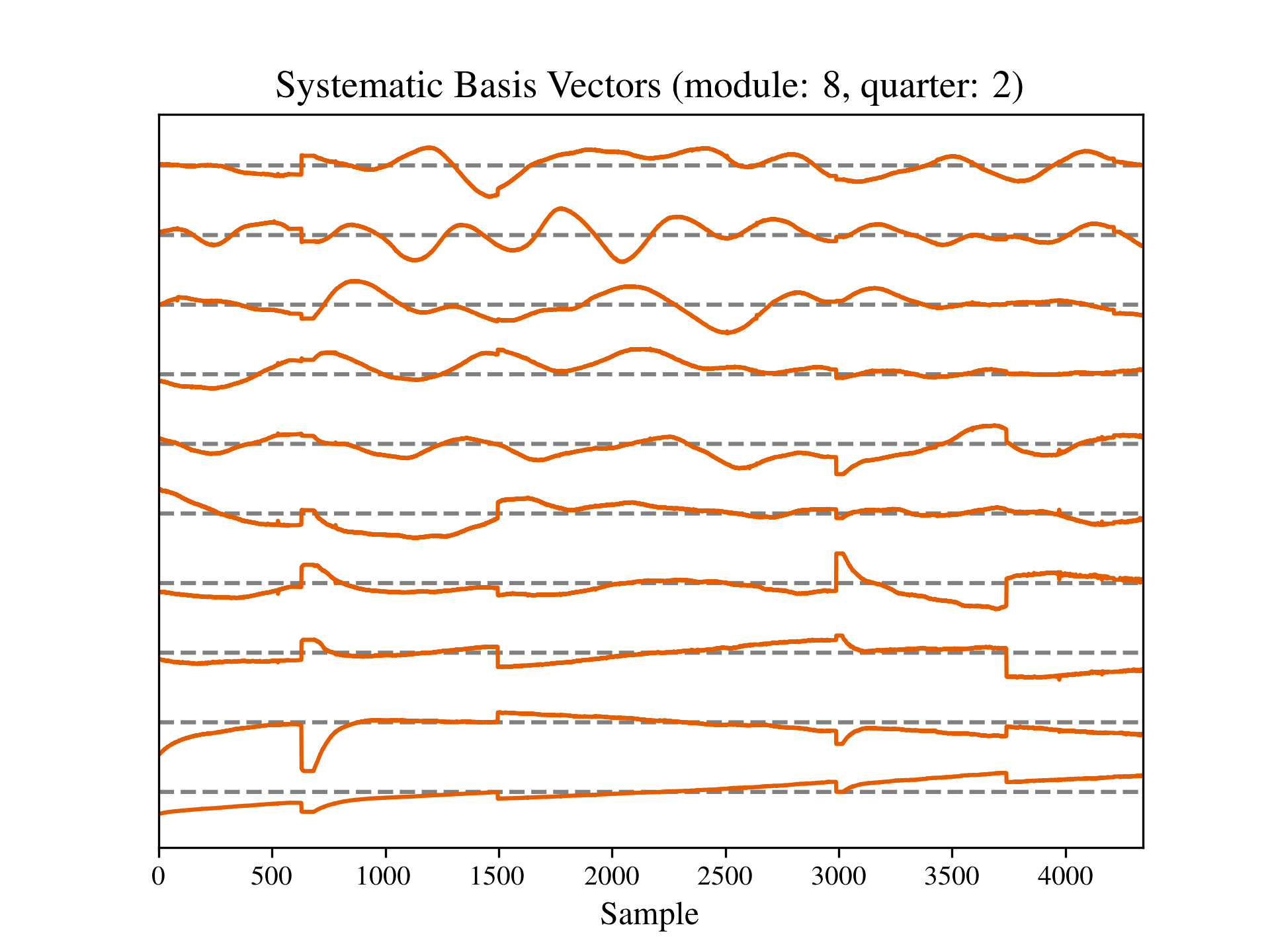}
\caption{A sample of 10 (out of model order $K=20$) systematic basis vectors $\{\mathbf{v}_k\}$ obtained using PCA for data from CCD module 8 during Kepler observing quarter 2. The basis vectors are ordered by principal value increasing from bottom to top. The x-axis denotes Kepler long-cadence time sample index ($\triangle t=29.4\ {\rm min}$). The y-axis is a uniform normalized zero-mean amplitude scale and is not annotated explicitly accordingly.
\label{fig:basisvec}}
\end{figure}

\begin{figure}[ht!]
\plotone{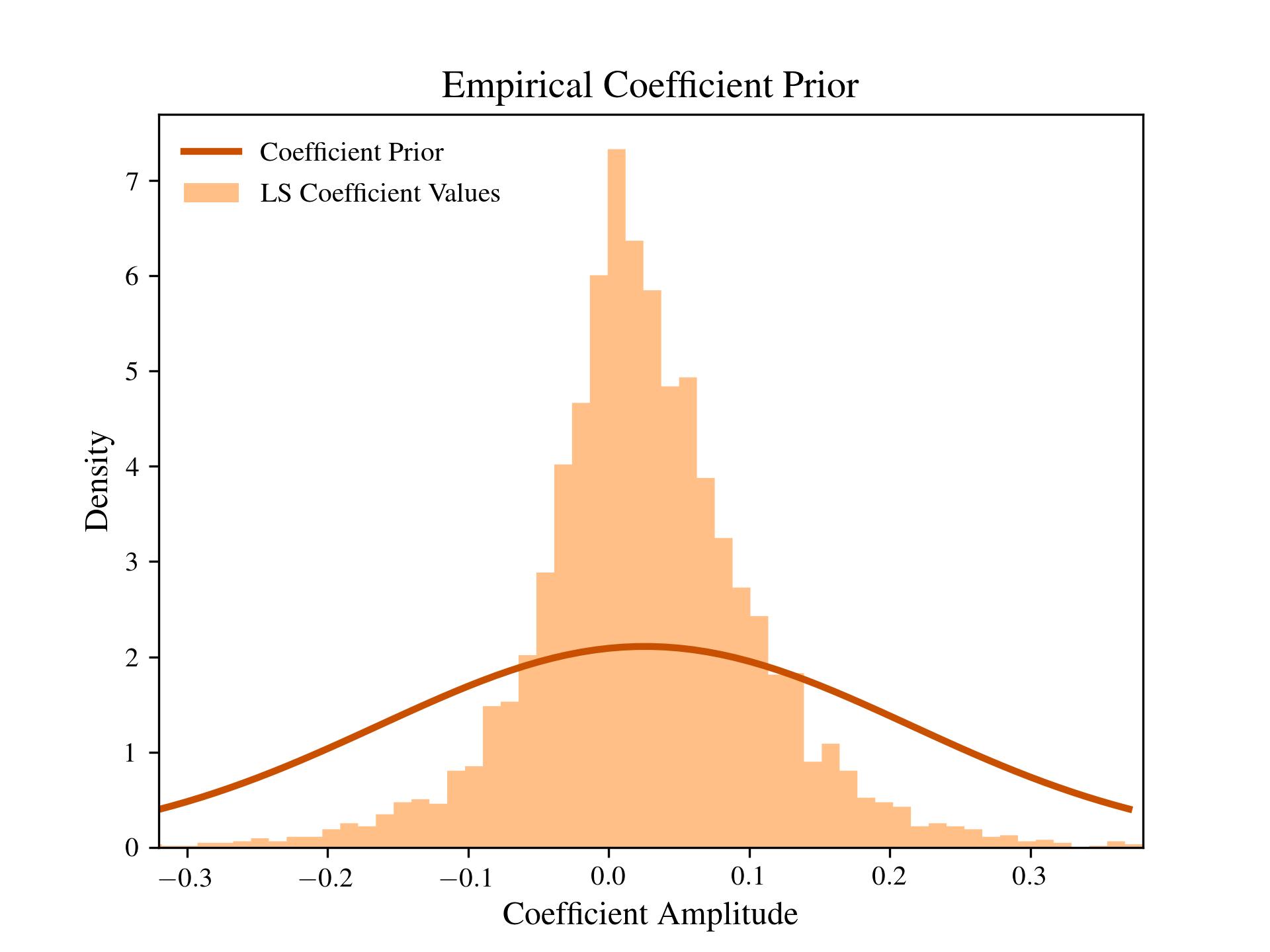}
\caption{The density of the least-square fit values for systematic noise basic vector coefficients ${\hat{c}}_i^1$ obtained for raw light curves $\{\mathbf{y}_i\}$  from CCD module 8 during Kepler observing quarter 2; superscript 1 denotes the first element of each coefficient vector. A Gaussian prior was fitted to the histogram and is shown as an overlay. The fitted prior is very broad due to a number of small outlier values as discussed in the text. The plots for other coefficient indexes (equivalently basis vector indexes) are similar in functional form. \label{fig:c1dist}}
\end{figure}

A systematic noise prior $p(\mathbf{c}_i)$ in Gaussian form  $\mathbf{c}_i \sim \mathcal{N}(\mathbf{\mu}_{c,i}, \Cov_{c,i})$ produces detectors in closed analytic form. A secondary motivation for adopting a Gaussian prior is that it has desirable non-parametric modelling properties when the underlying distribution is not fully known. Specifically, it is inherently regularizing and also allows correlations to be easily modelled. We motivate the choice of a Gaussian prior for $\mathbf{c}_i$ by examining the sample density of coefficient values $\{ \mathbf{\hat{c}}_i \}$ obtained via least-squares fits of $\{\mathbf{v}_k\}$ to $\{ \mathbf{y}_i \}$. Notwithstanding the issues with frequentist estimation of systematic noise noted earlier, it is assumed here that the least-squares fits $\{ \mathbf{\hat{c}}_i \}$ are approximately unbiased samples of the coefficient distribution of the population such that they may be used in inference of parameters describing $p(\mathbf{c}_i)$. Figure \ref{fig:basisvec} shows a subset of systematic basis noise vectors $\{ \mathbf{v}_k \}$ obtained here for CCD module 8 data during Kepler observing quarter 2 using PCA with model order $K=20$. This model order was chosen empirically as sufficient to be representative of the systematic noise whilst low enough as compared to the size of the light curve population $|I| = 5000$ so as to avoid over-fitting. In Figure \ref{fig:c1dist}, a histogram of the set of sample least-square coefficient values $\{ \hat{c}_i^1 \}$ is shown for the basis vectors in Figure \ref{fig:basisvec}; superscript 1 denotes the first element of each coefficient vector. A best-fit Gaussian coefficient prior is overlaid on the histogram. The histogram in Figure \ref{fig:c1dist} has approximately Gaussian form but with a narrower central mode and a broader tail in the distribution; collectively these broaden the overlaid Gaussian fit. The extremal outliers indicate excursions from our assumptions underlying the systematic noise model. We examine the impact of any non-Gaussianity in the systematic noise prior $p(\mathbf{c}_i)$ on detector performance in Section \ref{sec:discussion}.

\subsubsection{Detector A: Marginalization over Systematic Noise (Gaussian)}\label{sec:gauss_deta}

Detector A computes a LRT by marginalizing over the systematic noise distribution, as described in Section \ref{sec:deta}. A closed analytic form of this detector is derived here under the assumption of a Gaussian prior for the coefficients of the systematic noise basis vectors. If the coefficient prior $p(\mathbf{c}_i)$ is Gaussian $\mathbf{c}_i \thicksim N(\mu_{c,i},\Cov_{c,i})$, so too are the marginal probabilities of the observed signal $p_h(\mathbf{y}_i): h \in \{0, 1 \}$ (see Equation \ref{eq:detb_pzero} and Equation \ref{eq:detb_pone}):
\begin{align}
p_0(\mathbf{y})  \thicksim \mathcal{N} (\mathbf{V}\mu_{c,i},  \Cov_{\mathbf{s},i} + \mathbf{V} \Cov_{\mathbf{c},i} \mathbf{V^T})\\
p_1(\mathbf{y})  \thicksim \mathcal{N} (\mathbf{t} + \mathbf{V}\mu_{c,i},  \Cov_{\mathbf{s},i} + \mathbf{V} \Cov_{\mathbf{c},i} \mathbf{V^T})
\end{align}
where $\Cov_{\mathbf{s},i}$ is the covariance of the zero-mean stellar signal $\mathbf{s}_i$. The marginal detector formed from Equation \ref{eq:deta_lrt} is therefore equivalent under these assumption to detection of a known signal $\mathbf{t}$ in the presence of Gaussian noise $\mathbf{z}_i \thicksim \mathcal{N} (\mathbf{0} , \Cov_{\mathbf{s},i} + \mathbf{V} \Cov_{\mathbf{c},i} \mathbf{V^T})$ within the data $\mathbf{\hat{y}}_i = \mathbf{y}_i - \mathbf{V}\mu_{c,i}$. The matched filter is therefore optimal (Section \ref{sec:matched_filter}) with test statistic $T_i (\mathbf{\hat{y_i}})$:
\begin{align}
T_i (\mathbf{\hat{y_i}}) = \frac{\mathbf{\hat{y}}_i^T \mathbf{C_{z,i}}^{-1} \mathbf{t}}{ \sqrt{\mathbf{t}^T\mathbf{C_{z,i}}^{-1} \mathbf{t}}} \LRT{H_1}{H_0} \tau
\end{align}
\subsubsection{Detector B: Joint Transit and Systematic Noise Estimation (Gaussian)} \label{sec:detb_gauss}
Detector B produces MAP systematic estimates conditioned on the null and alternate hypothesis: $\mathbf{\hat{c}_{h,i}^{MAP}}: h \in \{0, 1 \}$ as defined in Equations \ref{eq:detb_c1map} and \ref{eq:detb_c0map}. These distinct conditional estimates are used as input to the detection test in Equation \ref{eq:detb}. Closed-form expressions for MAP/MMSE \footnote{Since both prior and likelihood are Gaussian, so too is the posterior as a Gaussian is a conjugate distribution \citep{wasserman_2013}. For a Gaussian distribution MAP and MMSE estimates are equivalent.} estimates $\mathbf{\hat{c}_{h,i}^{MAP/MMSE}}: h \in \{0, 1 \} $ are provided based on a Gaussian systematic noise prior $\mathbf{c}_i \thicksim N(\mu_{c,i},\Cov_{c,i})$ and a zero-mean stellar signal $\mathbf{s}_i$ with covariance $\Cov_{s,i}$. These estimates can be found either by computing the expectation with respect to the posterior distribution $\mathbb{E}_h(\mathbf{c_i}|\mathbf{y})$ or by maximizing the log of the posterior as shown in \citet{kep}:
\begin{align}
\mathbf{\hat{c}^{MAP/MMSE}_{0,i}} = (\mathbf{V^T}\Cov_{s,i}^{-1}\mathbf{V} + \Cov_{c,i}^{-1})^{-1} (\mathbf{V^T}\Cov_{s,i}^{-1}\mathbf{y}_i+  \Cov_{c,i}^{-1}\mu_{c,i})
\\
\mathbf{\hat{c}^{MAP/MMSE}_{1,i}} = (\mathbf{V^T}\Cov_{s,i}^{-1}\mathbf{V} + \Cov_{c,i}^{-1})^{-1} (\mathbf{V^T}\Cov_{s,i}^{-1}(\mathbf{y}_i - \mathbf{t})+  \Cov_{c,i}^{-1}\mu_{c,i})
\label{eq:map/mmse}
\end{align}

\subsubsection{Detector Parameter Estimation}
In their implementation, the joint detectors A and B described above require several parameters characterizing the systematic and stellar noise to be estimated or held fixed. This is not trivial as there are no clean, well-separated data. The parameters that need to be estimated include those of the Gaussian prior $\mathbf{c}_i \thicksim N(\mu_{c,i},\Cov_{c,i})$ for the coefficients $\mathbf{c}_i$ of the systematic noise basis vectors $\mathbf{v}_k : k \in K$, for fixed model order $K$. In addition, the parameters of the statistical distribution of the stellar signal $\mathbf{s}_i$ need to be estimated.

\begin{figure}[ht!]
\plotone{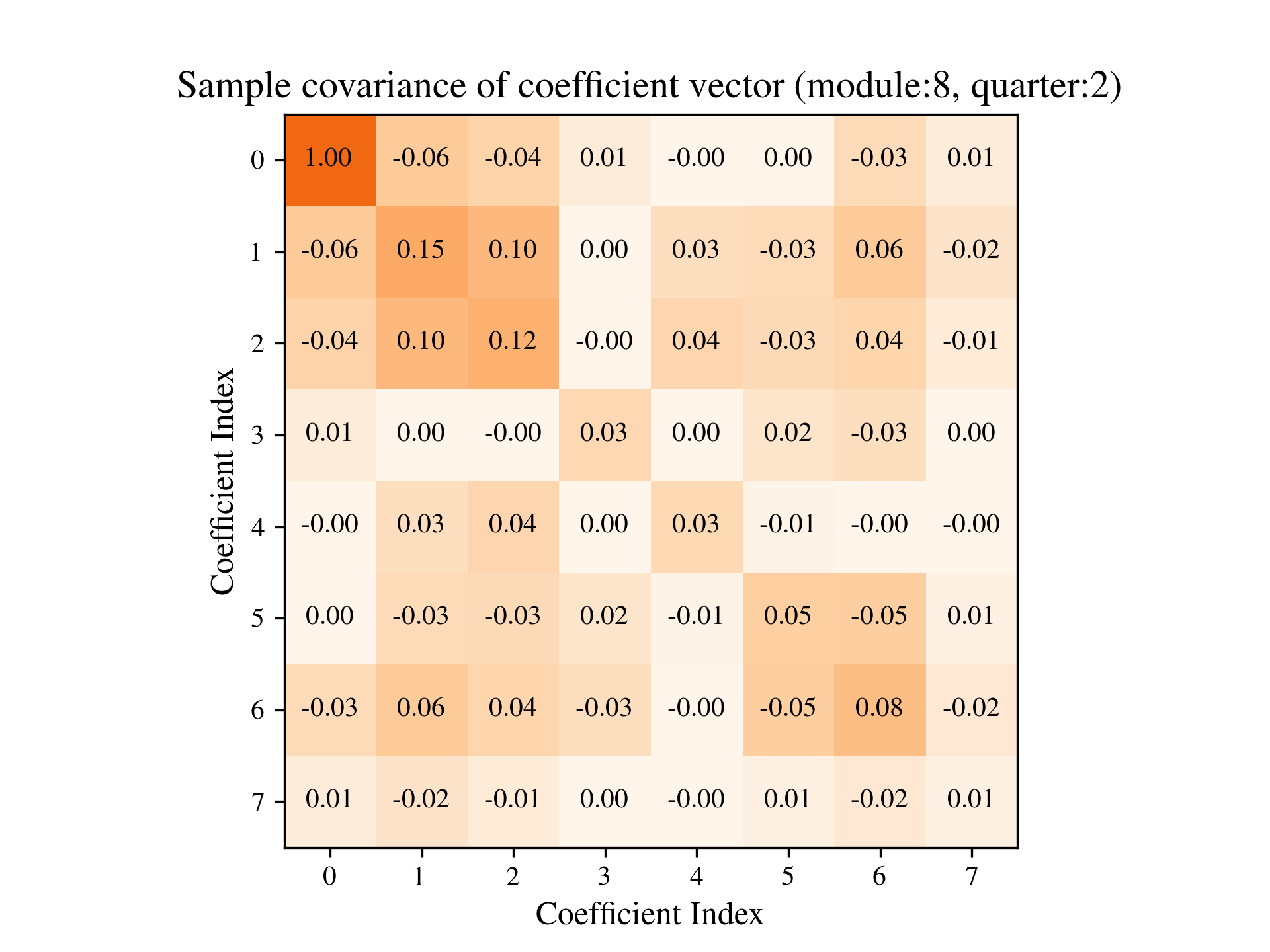}
\caption{Normalized sample covariance of coefficient value estimates $\{ \mathbf{\hat{c}}_i: i \in I \}$ obtained from least-square fits of $\{ \mathbf{v}_k : k \in K \}$ to light curves $\{ \mathbf{y}_i : i \in I \}$. These data are from the Kepler CCD module 8 during quarter 2. The
$ c_0 $ variance indicates that $\mathbf{v}_0$ is a dominant systematic trend across the light curve population. Correlations are also observed between basis signals. For ease of presentation, the covariance is only shown for 8 of 20 coefficient values and has been normalized by dividing the matrix by the maximum value.
\label{fig:samplecov}}
\end{figure}

In the current work we estimate the systematic noise basis vectors $ \{ \mathbf{v}_k \}$ using PCA as implemented in the Python module sklearn\footnote{https://scikit-learn.org/stable/}. To suppress the inclusion of stellar noise or dominating outlier lightcurves, the basis is constructed from $90 \% $ of the total lightcurves; those which have the lowest variance in absolute value. The parameter values of the Gaussian coefficient prior $\mathbf{c}_i \thicksim N(\mu_{c,i},\Cov_{c,i})$ are estimated directly from the set of coefficient estimates $\{ \mathbf{\hat{c}}_i \}$ obtained from least-squares fits of the systematic basis vectors $\{ \mathbf{v}_k \}$ to the raw light curves $\{\mathbf{y}_i\}$. Here we assume that for a population $I$ the same coeffient covariance may be used, denoted by $\Cov_{c, I}$. An example sample covariance is shown in Figure \ref{fig:samplecov}. It can  be seen that in this example the coefficient values are correlated (by the presence of non-zero off-diagonal elements). The PCA method finds an orthogonal set of basis vectors $\{ \mathbf{v}_k \}$, but there is no reason to expect independent systematic noise signals themselves to be orthogonal. Hence the orthogonalization procedure may distribute a systematic noise signal across multiple basis vectors. This will likely produce correlations between coefficients.

While the coefficient covariance $\Cov_{c, I}$ is estimated from the global population of least-square fits, the coefficient mean $\mu_{c,i}$ is taken to be the least-square coefficient vector $\mathbf{\hat{c}}_i$ obtained for the particular light curve under consideration.

As noted above, we assume that the stellar noise $\mathbf{s}_i$ is wide-sense-stationary (WSS) and zero-mean. This implies the stellar covariance matrix is Toeplitz and fully parameterizes the stellar noise. We therefore only need to estimate the stellar spectrum ($N$ free parameters) as opposed to a full covariance matrix ($N^2$ free parameters) The spectral estimation technique is not prescribed by the form of the detectors; here we use a smoothed periodogram \cite{kay} on the least-square cotrended light curves. 
\subsection{Numerical Simulations and Detector Performance Evaluation}
The joint detectors considered in this paper are defined in statistical and algorithmic terms in Section \ref{sec:jointdet}. Their specific design and implementation in the current work is described in Section \ref{sec:design_impl}. To evaluate the performance of the detectors we have conducted numerical single-transit injection tests using a subset of raw Kepler light curves from which confirmed exoplanet detections have been excluded. Confirmed exoplanets are those for which the initial detection has been validated by the project by a secondary analysis or follow-up observations. In addition, an initial feasibility test of the detectors against Kepler data containing known exoplanet detections was also performed. This Section describes the nature of these numerical studies augmented by a brief discussion of computational optimization and complexity issues relevant to these tests.

\subsubsection{Injection Tests}\label{sec:injtest}
\par We evaluate our detection performance using single-transit injection tests \citep{gilliland, weldrake2005absence, burke2006survey, burke2017planet} on raw Kepler simple aperture photometry (SAP) light curves \citep{jenkins2010overview} selected to exclude known exoplanet detections.  We use the long-cadence Kepler data in this analysis for which there is a  $29.4\ {\rm min}$ integration time \citep{noise}. A subset of $20,000$ such Kepler light curves were selected, comprising $5000$ light curves from each of the following Kepler CCD module and observing quarter pairs [M6:Q10, M8:Q2, M14:Q9, M18:Q4] \footnote{SAP flux light curves were downloaded using the Lightkurve library \citep{lk} }. These module and quarter pairs were selected randomly over time and CCD module position. Within each pair, the light curves were sorted by angular separation from the module reference point used by the MAST Kepler data archive and the first 5000 were selected from the sorted list. We exclude any light curves associated with exoplanets defined as confirmed by the NASA Exoplanet Archives\footnote{List of confirmed exoplanets from NASA Exoplanet Archives (https://exoplanetarchive.ipac.caltech.edu)}. A single synthetic transit is injected once per light curve, transit signals were simulated using the python transit\footnote{http://dfm.io/transit} library developed by D. Foreman-Mackey. These synthetic transit signals include limb-darkening \citep{mandel2002analytic, kip1} and a complete description of the transiting Keplerian orbital elements. The injected signals are drawn from a distribution of exoplanet population parameters given in Table \ref{tab:transitparams}. This population parameter distribution is informed by that used by \citet{foreman}, \citet{kip1}, and \citet{kip2} but it not identical. We adopted zero orbital eccentricity in the current work, amongst other changes.
\startlongtable\label{tab:transitparams}
\begin{deluxetable}{cc}
\tablecaption{Injected Signal Parameter Distribution} 
\tablehead{
\colhead{Transit Parameter} & \colhead{Distribution} 
}
\startdata
Period $P$ (days) & $U$(0.5, 40.) \\
Radius ratio of planet to host star (\%) & $U$(0.01, 0.2) \\
Transit epoch $t_0$ (days) & $U$(0, $P$) \\
Impact parameter (stellar radii) & $U$(0, 1) \\
Argument of periapse $\omega$ (rad) & $U$(-$\pi$, $\pi$)\\
Limb darkening parameters: $q_1$, $q_2$ & $U$(0, 1)
\enddata
\tablecomments{The distribution of injected signal parameters. A uniform probability density function over the domain $\{x_1,x_2\}$} is denoted as $U(x_1,x_2)$. The limb-darkening parameters are defined in \citet{kip1}; see also \citet{mandel2002analytic}. 
\end{deluxetable}

\subsubsection{Standard Model Processing}\label{sec:stdmodel}
As a reference detector we adopt the standard heuristic of sequential cotrending followed by detection \citep{Stumpe_2012}. We term this the standard model in what follows and provide our own implementation of this detector in the current work. In the standard model the cotrending is performed assuming the light curve contains no transit signal: $\mathbf{y}_i = \mathbf{s}_i + \mathbf{V}\mathbf{c}_i$, analogous to hypothesis $H_0$ (Equation \ref{eq:detb_h0}). A MAP/MMSE estimator for the systematic noise is constructed equivalent to $\mathbf{\hat{c}_0^{MAP/MMSE}}$ (Equation \ref{eq:detb_c0map}) and applying the same Bayesian priors for systematic noise and stellar noise as used by detectors A and B. Detection is then performed on the cotrended light curves $\mathbf{\hat{y}}_{i} = \mathbf{y}_i - \mathbf{V} \mathbf{\hat{c}_{i,0}^{MAP/MMSE}}$ using the matched filter in Equation (\ref{eq:matchedfilter}) with $\Cov_n = \Cov_{s,i}$.

\subsubsection{Transit Search Space Optimization}\label{sec:search_opt}
\par As discussed in section \ref{sec:transit}, transit detection requires testing every candidate transit signal $\mathbf{t} \in \mathbf{T}$ to find that which maximizes the test statistic $T(\mathbf{y})$. The transit space $\mathbf{T}$ is typically populated by periodic box functions over a range of candidate orbital periods $P$, transit durations $d$, and epoch times $t_0$ in the functional form described by Equation ~\ref{eq:transit}. Transit depth $\alpha$ is omitted here as our detector forms generally do not make use of this parameter. The dimensionality of the transit parameter search space is therefore intrinsically large and the transit detection problem computationally expensive. This computational cost can be reduced sharply by constraining the range of epochs $\{ t_0 \}$ for a candidate transit with a certain period and duration using the method of phase correlation \citep{phase}. We adopt this method in our numerical studies due to the significant reduction in computational cost. The phase correlation method and its application to epoch estimation is described in Appendix ~\ref{sec:phasecorr}.
\par The phase-correlation method, however, requires the use of cotrended light curves for sufficient accuracy in the estimated epochs $\{t_0\}$. This raises the concern that the transit signal may not be detected optimally due to the use of the cotrended as opposed to raw light curves. To verify that this approach does not decrease detection efficiency we compared detection results with and without phase correlation using single-transit injection tests over 5000 light curves from the broader injection test data described in Section ~\ref{sec:injtest} selected here from [M8:Q2]. We define detection efficiency in this context as the rate of correct detection of the known injected signals as described in Section ~\ref{sec:detperformance}. In each case the standard reference detector defined in Section ~\ref{sec:stdmodel} was used; as described above this detector comprises sequential cotrending and detection steps. In the test without phase correlation, a 3-dimensional transit signal parameter search space was used as defined in Table ~\ref{tab:search_space}; the standard detector operated on the raw light curves over this gridded search space. In the phase-correlation test, a transit epoch $t_0$ was estimated from each least-square cotrended light curve using the phase-correlation method. The standard detector was then applied to the raw light curves holding $\{t_0\}$ fixed to the phase-correlation estimate but searching over a residual 2-dimensional search space in period and duration as tabulated in Table ~\ref{tab:search_space}. 
\startlongtable
\begin{deluxetable}{cccc}
\tablecaption{Correlation Verification Transit Search Space \label{tab:search_space}}
\tablehead{
\colhead{Transit Parameter} & \colhead{Range} & \colhead{Step Size} &
\colhead{Physical Units}\\
\colhead{} & \colhead{($\triangle t_{LC}$)} & \colhead{($\triangle t_{LC}$)} & 
}
\startdata
Period  & $P_{LC} \in [20, 2125]$ & 1 & $P \in [1, 43.4]$ d \\
Duration & $d_{LC} \in [3, 11]$ & 2 & $d \in [1.4, 5.4]$ h \\
Epoch  & [0, $P_{LC}$] & $\frac{d_{LC}}{2}$ & 
\enddata
\tablecomments{The long cadence sample integration time is $\triangle t_{LC}=29.4 \textrm{min}$}
\end{deluxetable}

\begin{figure}[ht!]
\plotone{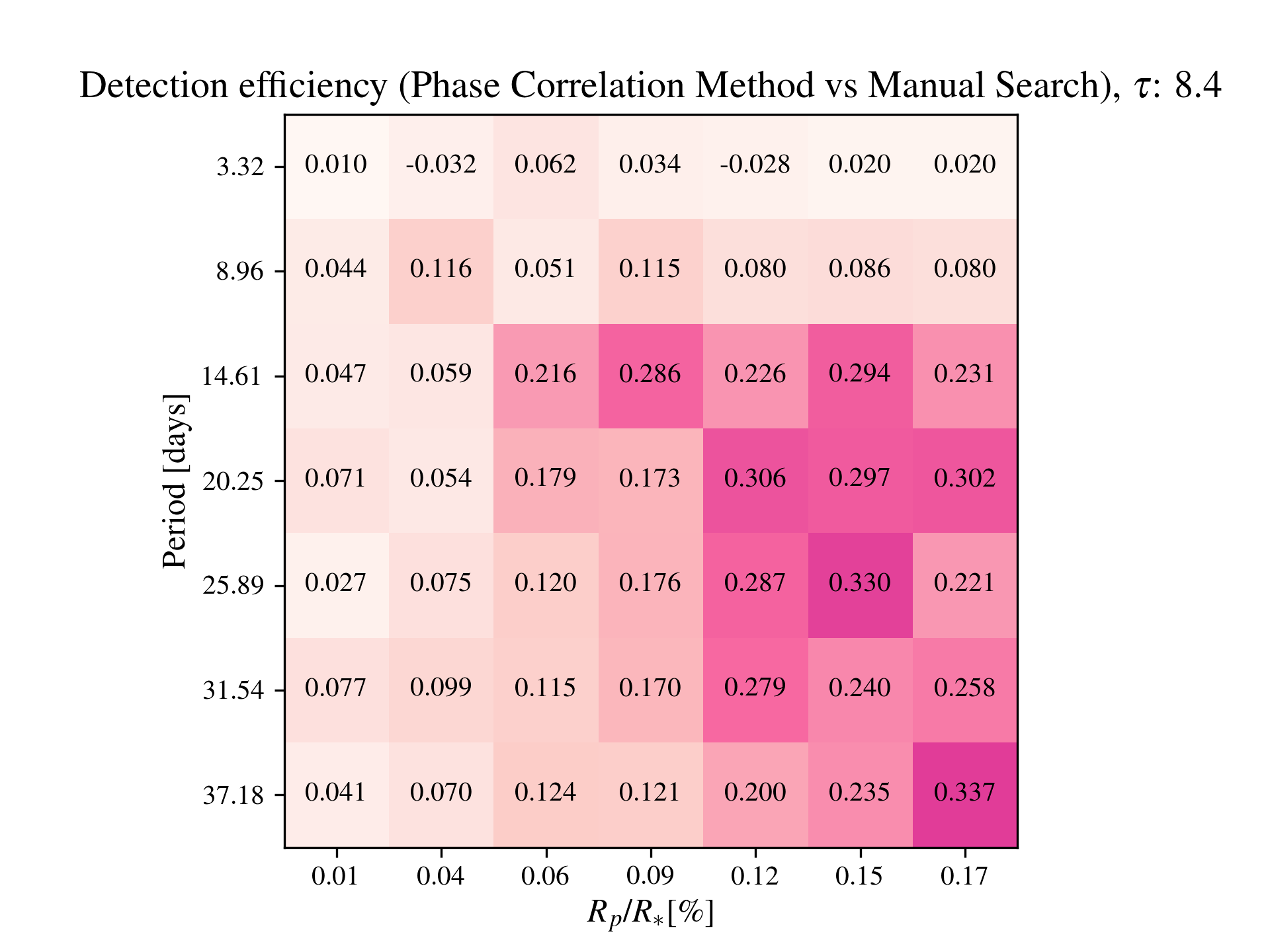}
\caption{Detection efficiency improvement of the phase correlation method as compared to a direct search is shown for a threshold of $\tau = 8.4 $ broken down by injected transit orbital period and transit depth. This data is from the phase correlation injection verification tests described in Section \ref{sec:search_opt}. The phase correlation method demonstrated no marked loss in detection efficiency for weak signals. }
\label{fig:corr_eff}
\end{figure}

The test data here comprise actual raw light curves from which confirmed exoplanets have been excluded; however, the data cannot be shown provably to exclude any hitherto undetected transit signals. As such, we define a quasi false-alarm rate as the rate of incorrect detection with respect to the injected signal set. The detection tests using the phase-correlation method show an improvement in detection efficiency of $14\%$ and a reduction in quasi false-alarm rate of $16\%$ (at a detection threshold $\tau = 8.4$) over the detection tests for which a direct search was performed. A comparison of detection efficiency broken down by transit parameter values is shown in Figure ~\ref{fig:corr_eff} and demonstrates no marked reduction in detection for weak signals. The phase correlation method described in Appendix \ref{sec:phasecorr} by definition has a maximum accuracy $\triangle t_0$ in epoch $t_0$ of \textbf{one long-cadence sample $\triangle t_{LC}$ (a single 'pixel')}. Non-additive noise will reduce the accuracy of the method. By considering the minimum required correlation between a measured transit and a parametrized transit model $(t_0,d,P)$, \citet{jenkins_2010} provide an analysis motivating a default search spacing in epoch of $\triangle t_0=\frac{d}{10}$ for the Transiting Planet Search (TPS) module in the Kepler science pipeline. Our direct search here used a step size $\triangle t_0=\frac{d}{2}$ (see Table \ref{tab:search_space}) which is sub-optimal relative to the maximum epoch accuracy of the phase correlation method thereby possibly explaining the improved detection efficiency of the latter method here. We stress here that these tests only demonstrate that the phase correlation optimization is suitable for the transit search space considered here; generalization to broader applicability is left to future work.

The ratio of computational cost between the test without phase-correlation and that using phase correlation was $\sim 10^2:1$. As discussed further in Section \ref{sec:discussion} the freed computational resources allow more refined searches in the remaining transit parameters and can be argued to improve overall accuracy in that sense. As a result of the positive outcome of this verification test and the significant associated reduction in computational cost, we used the phase-correlation method described in \textbf{Appendix} ~\ref{sec:phasecorr} to estimate transit epochs $\{t_0\}$ in our full injection tests described in Section ~\ref{sec:injtest}.
For the full injection tests  the transit signal parameter search space is informed broadly by \citet{jenkins_2002}. However, given our use of the phase correlation method and the associated freed computational resources, we search (in units of long-cadence samples $\triangle t_{LC}=29.4\ {\rm min}$) over the period range [20, 2125] with a .25 step size and within the following set of transit durations (in long-cadence samples): [2,3,4,5,6,7,9,10,12].
Detector B additionally requires a search over transit depth parameter $\alpha$. This arises from the estimation of transit-dependent systematics as in Equation \ref{eq:map/mmse} for $\mathbf{\hat{c}_{1,i}^{MAP/MMSE}}$. In contrast the matched filter depends purely on the shape of a transit signal (defined by $t_0$, $d$, $P$) and not the signal strength $\alpha$. One can see this property by considering a scaled signal $\alpha \mathbf{t}$ in the matched filter function \ref{eq:matchedfilter}; the scaling $\alpha$ immediately cancels. For the rest of the detectors, the only step that is dependent on a transit signal is a matched filter step, ergo they do not depend on a transit depth parameter. For detector B we search over four equally-spaced transit depths $\alpha \in \{0.2, 0.5, 0.8, 1.1\}$, scaled by the maximum range of the least-square cotrended light curve under consideration. Our choice of transit depth sampling is exploratory but proved practical. We note however that it sets a limit on the detectability of signals with transit depths below 20$\%$ of the cotrended lightcurve. Future work will consider optimized sampling schemes for transit depth including estimated noise levels.

\subsubsection{Computational complexity}\label{sec:complexity}
\par The computational complexity of the detectors and their key constituent operations acting on a single light curve is summarized in Table \ref{tab:table3}. The computational complexities are expressed in terms of the length of the light curve $N$ and the size of the transit signal search space $|\mathbf{T}|$ as defined in the introduction of Section \ref{sec:detmodel}. These complexities are  specific to the case of a Gaussian systematics prior described in Section \ref{sec:gaussian_sys}.

A time-domain matched filter defined in the form of Equation \ref{eq:matchedfilter} has computational complexity dominated by the product of a $[N \times N]$ matrix and a length $[N]$ vector, therefore its complexity is $O(N^2)$. A matched filter implemented in the Fourier domain involves computing Fast Fourier Transforms (FFT) and inner products of length $N$ vectors \citep{kay}. Between these operations the FFT is more computationally intensive, hence the Fourier-domain matched filter is  $O(N\log N)$ \citep{bracewell2000fourier}. 

The computational complexity of MAP/MMSE systematics estimation is defined by the form of Equation \ref{eq:map/mmse}. For a particular light curve we can reuse many of the computed terms during multiple transit tests. Considering transit-dependent terms, the dominant computational term is the product of the $[N \times N]$ matrix $\Cov_{s,i}^{-1}$ and the length $N$ vector $\mathbf{t}$; consequentially this computation is $O(N^2)$. Once per light curve, a matrix inversion of the $N \times N$ matrix $\Cov_{s,i}$ is performed and this operation is $O(N^3)$. However it is not leading order since generally $|\mathbf{T}|N^2 \gg N^3$. As such, MAP/MMSE systematics estimation has computational complexity $O(N^2)$ per light curve per transit.

The detector complexities are determined by the form of the matched filter used and scaled by the size of the transit search space. The standard detector (Section \ref{sec:stdmodel}) is the most computationally efficient detection strategy as per transit the only computation performed is a Fourier-domain matched filter. The standard detector searches a transit space of size $|\mathbf{T}|$ and therefore has a computational complexity $O(|\mathbf{T}|N \log N)$. 

Detector A (Section \ref{sec:gauss_deta}) searches for a transit signal contained in non-WSS Gaussian noise, therefore a time-domain matched filter must be used for each transit test. The search space is of size $|\mathbf{T}|$ and the net computational complexity is $O(|\mathbf{T}|N^2)$.

Detector B (Section \ref{sec:detb_gauss}) uses a larger search space than the other detection strategies as it includes transit depth $\alpha$; this search space was described in Section \ref{sec:search_opt} and is of size $|T^*|$. In addition to this increased search space, this detector must compute a MAP/MMSE systematics estimate once per transit which is then used as input into a Fourier-domain matched filter. The net computational complexity is therefore $O(|\mathbf{T}^*|N^2)$.

\startlongtable
\begin{deluxetable}{cc}
\tablecaption{Computational complexity of operations on a single light curve of length $N$ \label{tab:table3}}
\tablehead{
\colhead{Operation} & \colhead{Complexity} 
}
\startdata
Time-domain matched filter  &  $O(N^2)$ \\
Fourier-domain matched filter & $O(N \log N)$ \\
MAP/MMSE systematics estimation & $O(N^2)$ \\
Standard detector (Gaussian prior) & $O(|\mathbf{T}| N \log N)$ \\
Detector A (Gaussian prior) & $O(|\mathbf{T}| N^2)$ \\
Detector B (Gaussian prior) & $O(|\mathbf{T}^*| N^2)$
\enddata
\tablecomments{Where $|\mathbf{T}|$ is the search space size. For detector B the search space size $|\mathbf{T}^*|$ is generally larger as one must additionally search over candidate transit depths $|\alpha|$.}
\end{deluxetable}

Approximate elapsed wall-clock run times are summarized in Table \ref{tab:runtimes}. Transit detection tests were parallelized with one lightcurve assigned to each CPU core. All runs were performed on the Blue Waters petascale system at UIUC/NCSA \citep{Bode_2013}. This is a Cray XE/XK system with a peak performance of 13.34 PF\footnote{https://bluewaters.ncsa.illinois.edu/hardware-summary}.

\startlongtable
\begin{deluxetable}{ccc}
\tablecaption{Average single core run times per light curve transit search \label{tab:runtimes}}
\tablehead{
\colhead{Detector} & \colhead{Run time per lightcurve}  & \colhead{Run time per lightcurve per transit} \\
\colhead{} & \colhead{(hr)} & \colhead{(s)}
}
\startdata
Standard detector & 1.5 & 0.06 \\
Detector A & 5 & 0.18 \\
Detector B & 30 & 0.28
\enddata
\tablecomments{The transit search space sizes for these runs are $|\mathbf{T}| \approx 10^5$ and $|\mathbf{T}^*| \approx 4\times 10^5$. All detectors used a Gaussian prior. All run times are approximate elapsed wall-clock run times.}
\end{deluxetable}

\subsubsection{Feasibility Test: Kepler Data containing Exoplanets}\label{sec:kepler_data}
We conducted an initial feasibility test using detectors A, B and the standard detector on a subset of Kepler light curves that did not exclude known exoplanets. These tests were designed to demonstrate initial detection feasibility only on real exoplanet transit signatures.  Detection tests were performed over the transit search space identical to that used in injection tests. The transit signal parameter search space is described in Section ~\ref{sec:search_opt}.  A total of 2000 light curves were analysed in this test, 1000 light curves were selected from CCD module 2 over observing quarters [Q2, Q10, Q14] and and additional 1000 light curves were selected from CCD module 12 over observing quarters [Q3, Q7, Q15]. These CCD modules and observing quarters were chosen randomly over time and across CCD module. As was performed for the injection tests, the light curves for each module were first sorted by angular separation from the module reference point used by the MAST Kepler data archive before the first 1000 were selected. No explicit selection for CCD module output was applied: module 2 data included outputs 3 and 4 while module 12 data included only output 4. For our detection tests we did not use stitched quarters but instead performed separate detection tests on each of the quarters and computed an averaged test statistic (per transit over time).
\section{Results}\label{sec:results}

\subsection{Detector Performance: Injection Tests}\label{sec:detperformance}
\par Detection efficiency and quasi false-alarm rate are defined in the context of the recovery of injected signals in Section \ref{sec:search_opt}.
A detection occurs whenever there is a test statistic above the detection threshold $\tau$.

We consider a correct detection of an injection signal to occur if the maximum test statistic above threshold satisfies both of the following requirements: i) The detected orbital period is within 3 hours of the true injected orbital period; and ii) For an injected transit signal $\mathbf{t_r}$ and detected transit signal $\mathbf{t_d}$, the cosine similarity satisifies the condition $\frac{\mathbf{t_d}^T \mathbf{t_r}}{|\mathbf{t_r}||\mathbf{t_d}|} > \frac{1}{2}$. This threshold ensures that for an injected and detected transit of identical duration, the error in estimated epoch does not exceed half the transit duration. We note that the limb-darkened injected transit has a different functional form from the detected periodic box transit function; the correlation value will therefore be slightly lower than expected for a correct match.

Since the purpose of these tests is a comparison of detection strategies we do not seek to stringently reduce the false-alarm rate and thus require only two transit events (passes of an exoplanet) for a detection as opposed to the standard three transits \citep{burke2017planet}.

\begin{figure}[ht!]
\plotone{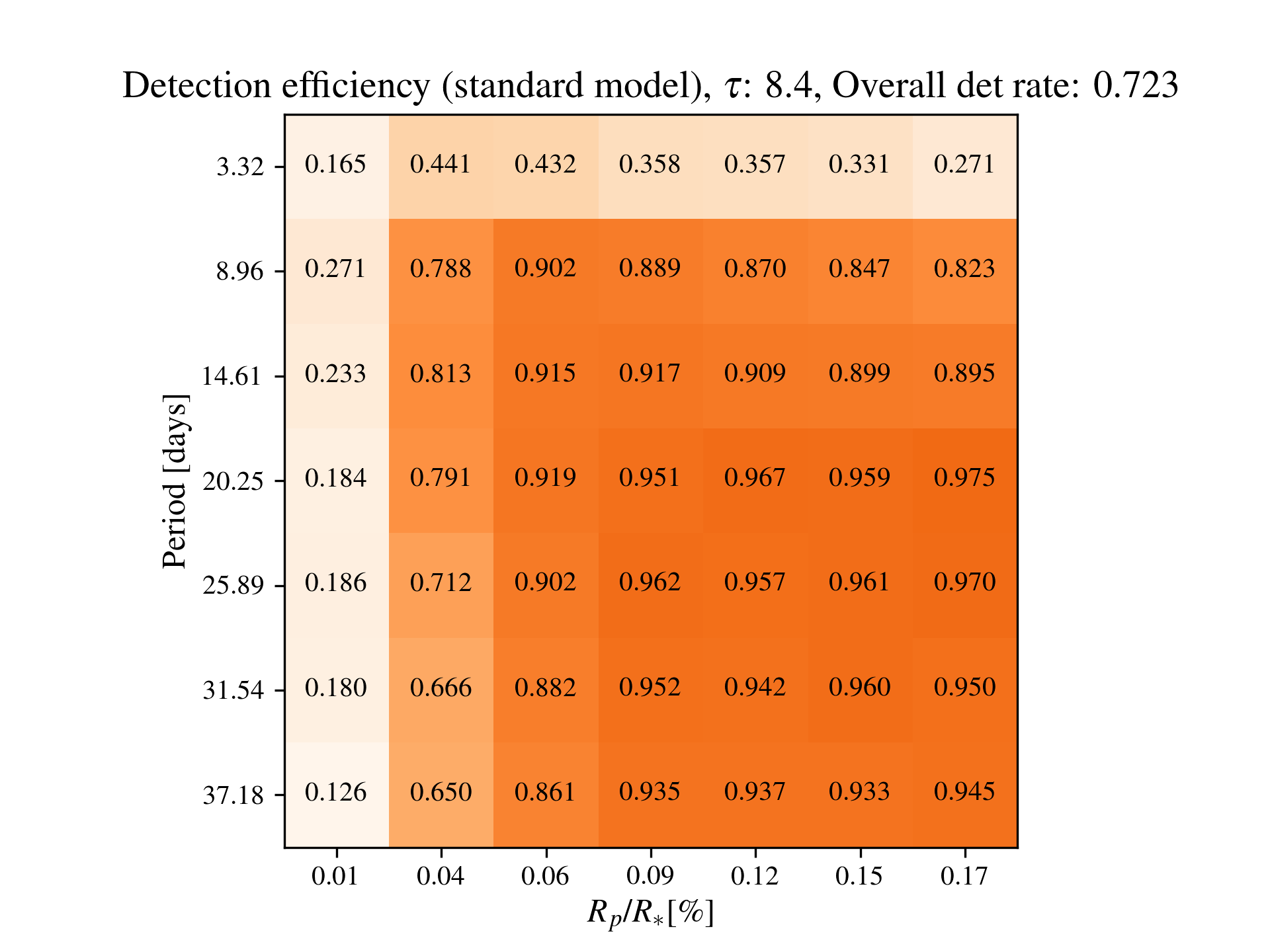}
\caption{The standard model detection efficiency for a threshold of $\tau = 8.4 $ computed using injection tests. The range of synthetic transit signal parameters can be found in Section \ref{tab:transitparams}.}
\label{fig:standdetect}
\end{figure}

\begin{figure}[ht!]
\plotone{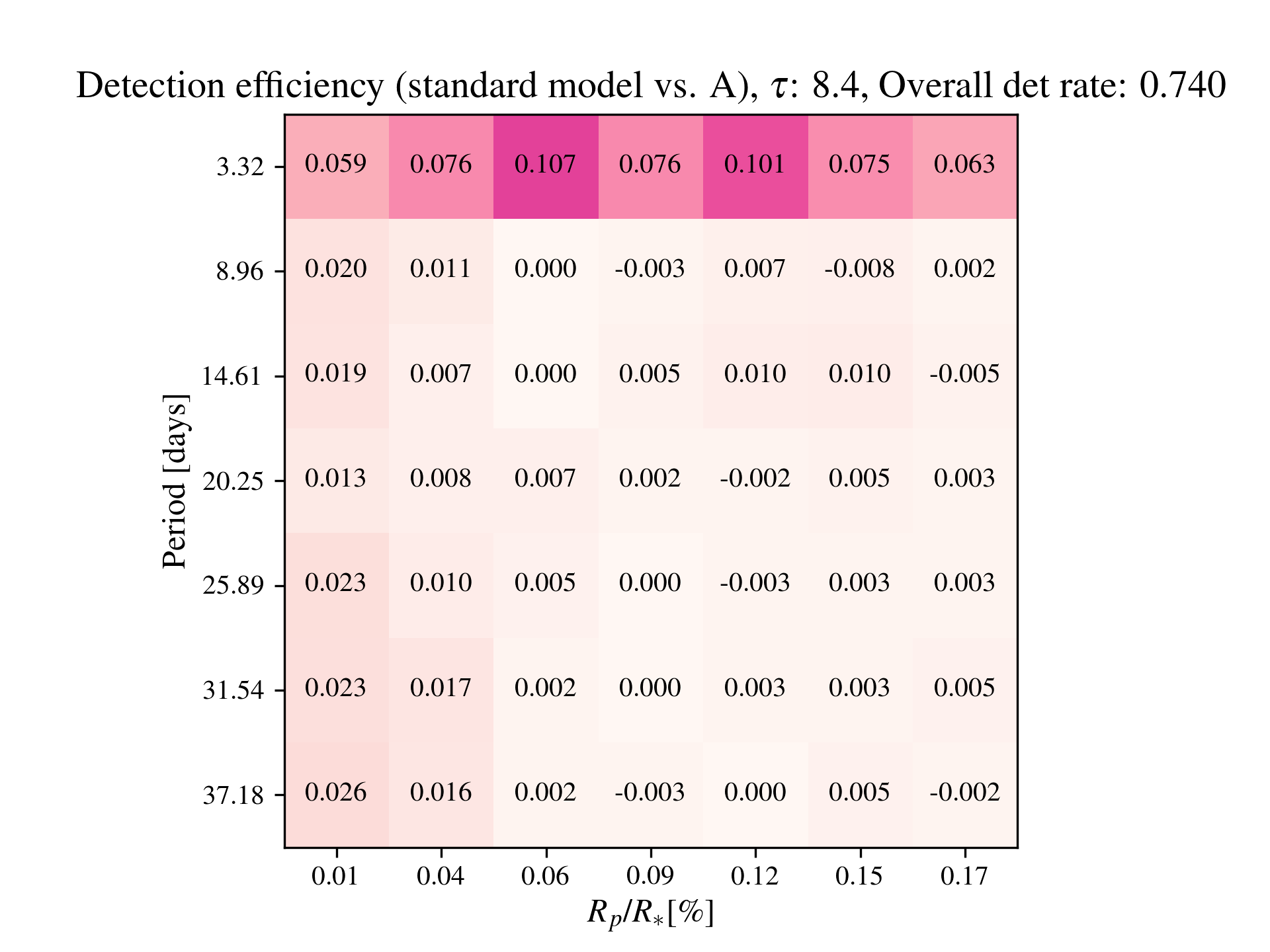}
\caption{Detector A detection efficiency compared to the standard detector for a threshold of $\tau = 8.4 $ computed using injection tests. The range of synthetic transit signal parameters can be found in Section \ref{tab:transitparams}.}
\label{fig:deta}

\end{figure}

\begin{figure}[ht!]
\plotone{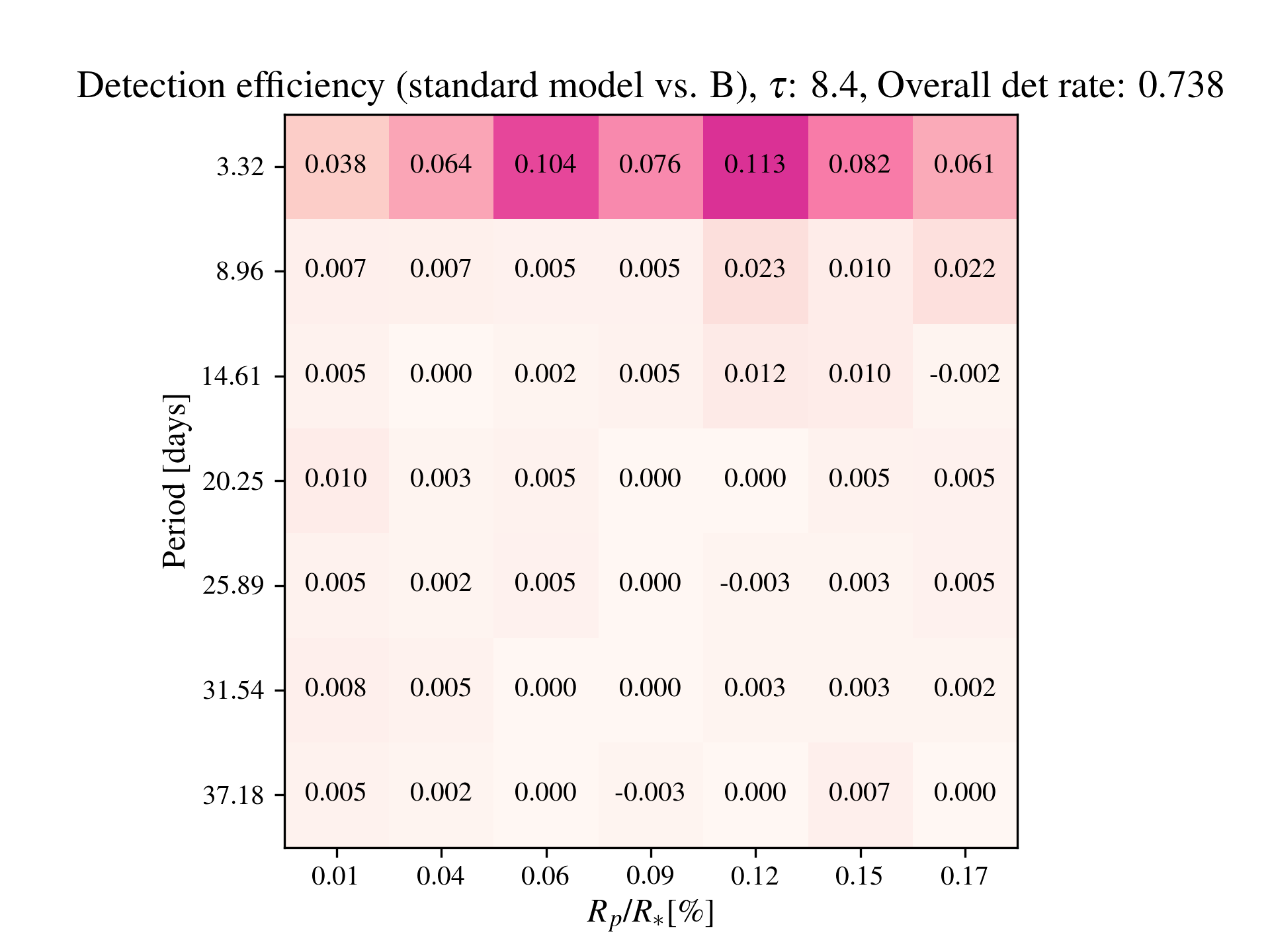}
\caption{Detector B detection efficiency compared to the standard detector for a threshold of $\tau = 8.4 $ computed using injection tests. The range of synthetic transit signal parameters can be found in 
Section \ref{tab:transitparams}.}
\label{fig:detb}
\end{figure}

When comparing Neyman-Pearson detectors, a detector is considered optimal if its detection rate is maximized for a fixed rate of false alarm \citep{kaybook, wasserman_2013}. We adopt a detection threshold $\tau = 8.4$ in comparing detection efficiency across the detectors considered here as it achieved a consistent quasi-false-alarm rate for these detectors of $13\pm 1 \%$. The detection efficiency broken down by orbital period $P$ and radius of planet-to-star ratio $\frac{R_p}{R_*}$ of the injected transit are show in Figure \ref{fig:standdetect} for the standard model. For the remaining detectors we display the difference in detection efficiency relative to the standard model. This is depicted in Figure \ref{fig:deta} for detector A and Figure \ref{fig:detb} for detector B.

\begin{figure}[ht!]
\plotone{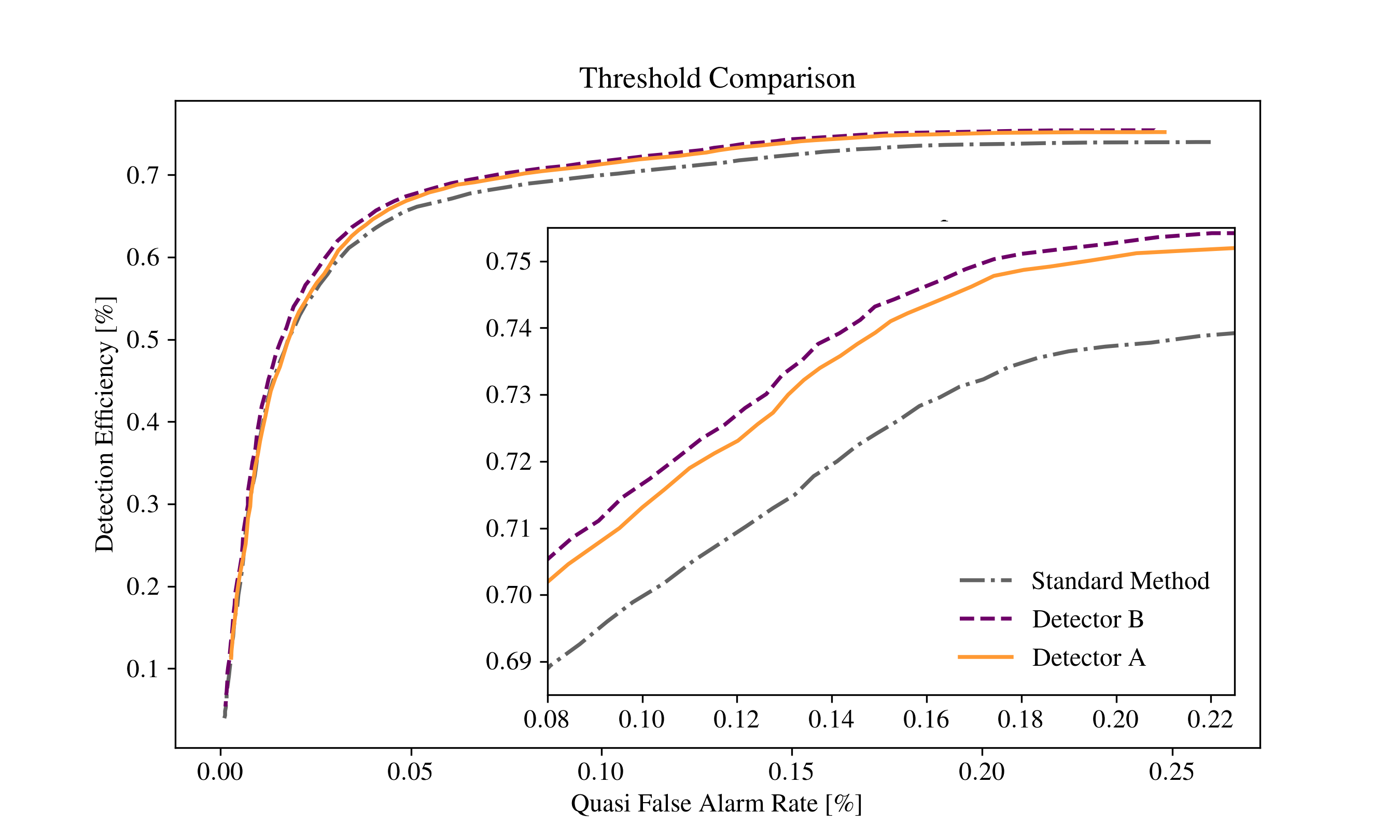}
\caption{A plot of detection efficiency against quasi-false-alarm rate in injection tests for decreasing threshold. Joint Bayesian detectors A and B on average produced a $\sim2\%$ increase in detection efficiency for the same quasi-false-alarm rate. The inset shows a magnified view of a region of the enclosing outer plot.}
\label{fig:ROC}
\end{figure}

The detection rate as a function of quasi-false-alarm rate for the detectors is plotted in Figure \ref{fig:ROC}. 

\subsection{Kepler Data Detections}\label{sec:kepler_results}
As described in Section \ref{sec:kepler_data} an initial feasibility test was conducted using these detectors on a subset of 2000 Kepler light curves from which prior exoplanet detections were not excluded; this subset contained 17 confirmed exoplanets. Performance was measured at a detection threshold of $\tau=7.5$, a lower detection threshold was used relative to the injection test value $\tau=8.4$ given expected suboptimal performance on real transit data. Detector A produced $252$ detections, $9/17$ of which are confirmed exoplanets and $76$ are threshold crossing events (TCE) \citep{jenkins2010overview}. Detector B produced $219$ detections, $9/17$ of which are confirmed exoplanets and $67$ are TCEs. The standard detector produced $187$ detections, $8/17$ of which are confirmed exoplanets and $59$ are TCEs.

\begin{figure}[ht!]
\plotone{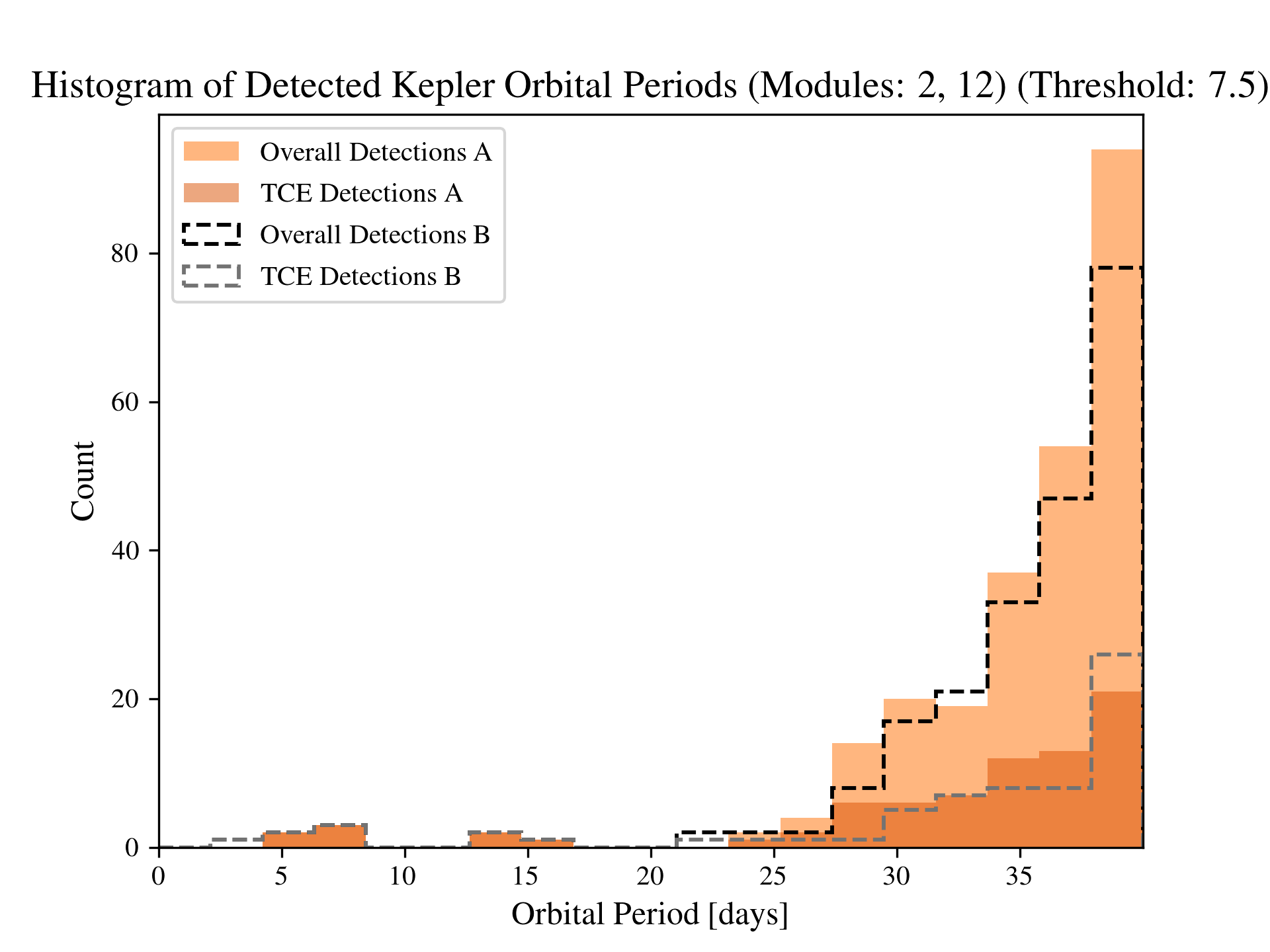}
\caption{Histogram of detected Kepler orbital periods from a sample of 2000 light curves selected from CCD modules 2 and 12. Results for detector's A and B are shown as indicated. Those detections which are also threshold crossing events (TCE's) are marked.}
\label{fig:histp}
\end{figure}

On visual inspection we find no new convincing exoplanet candidates in the complete set of detections. A histogram of the detected orbital periods is shown in Figure \ref{fig:histp}, in which detections that are also TCEs are marked.

We emphasize that this initial feasibility test on Kepler data containing exoplanets is not intended nor designed as a comparison of the statistical performance of these exploratory detectors against the Kepler science pipeline. The Kepler results are from a full multi-quarter analysis and are used here only as a test of the initial feasibility of our detectors in recovering known exoplanets and demonstrating consistent results.

\section{Discussion}\label{sec:discussion}

\par As shown in Figures \ref{fig:standdetect} to \ref{fig:detb}, in injection tests with a fixed threshold, the joint Bayesian detectors A and B (Section \ref{sec:jointdet}) achieve an overall detection efficiency improvement of $\sim 2\%$ over the reference standard processing model (Section \ref{sec:stdmodel}). As noted above, the relative detection efficiencies were assessed at a comparable quasi-false-alarm rate for a fixed threshold. 

As shown in Figure \ref{fig:ROC} for a fixed quasi-false-alarm rate, injection tests show that the $\sim 2\%$ improvement in detection efficiency for detectors A and B relative to the standard model remains consistent for every quasi-false-alarm rate above $4\%$. Furthermore below this rate the detectors A and B continue to outperform the standard model; this suggests that the joint detection strategies are Neyman-Pearson optimal \citep{wasserman_2013}.  As defined in Section \ref{sec:search_opt}, the quasi-false-alarm rate may overestimate the true false-alarm rate due to the presence of hitherto unknown actual detections in the Kepler data used for the injection tests despite the exclusion of confirmed exoplanet detections from these data. The quasi false-alarm rate can however be argued as a reasonable proxy for the true false-alarm rate. Specifically, a false injection test detection does not require that the true injected transit signal be below the detection threshold, only that another transit signal produce a stronger test statistic. This condition can be expected to have low probability however, given that prior undetected transit signals are likely to be weaker than the injected signals in general. In addition, although the quasi-false-alarm rate may overestimate the true false-alarm rate, we expect it to do so monotonically, lending validity to its use as a proxy in comparative detector studies.

Figure \ref{fig:ROC} shows that detector B marginally outperforms detector A. We speculate that this may be partially explained by deviations from Gaussianity between the sample and fitted systematic noise coefficient prior $p(\mathbf{c}_i)$ discussed in Section \ref{sec:gaussian_sys}. In a direct systematics estimate (detector B) a broader prior $p(\mathbf{c}_i)$ simply allows more variability about the mean to obtain a maximizing estimate under each hypothesis model. In a marginalization scheme (detector A) this may lead to weakened test statistics by including likelihoods for a number of improbable systematics estimates. In future work we will explore methods to constrain the systematics prior to more closely approximate the central mode (see Figure \ref{fig:c1dist}).

\par Much of the improvement in detection efficiency for detectors A and B is concentrated in exoplanets with shorter orbital periods ($P < 10 {\rm d}$ ) or with low ratios of planet to host star radii $\frac{R_p}{R_*}$ ($<0.05\%$). Broadly therefore, the improvements occur in short-period, low transit-depth populations within our sample. In an analysis of the detection efficiency of the Kepler pipeline \citep{Christiansen_2013, Christiansen_2015}, a drop off in detectibility was shown for exoplanets with orbital periods $P < 3 {\rm d}$. The authors demonstrate that the process of removing harmonics (residual high-frequency stellar noise left over after PDC) \citep{jenkins, jenkins2010overview} prior to transit detection may distort short period transit signals. The distribution of detection efficiency improvement as a function of orbital period for the detectors in the current work suggests that joint modeling of the systematic noise and transit signal mitigates this effect as it is better able to jointly differentiate between high frequency noise and short period transit signals. Specifically, we propose that a joint modeling approach, though computationally expensive, may be particularly effective when probing the aforementioned exoplanet populations. In general, it may be fruitful to use adaptive detection strategies in different parts of the transit parameter search space and in different SNR regimes. Further investigation of this effect over larger data samples is required. As noted earlier, \citet{foreman} have implemented a non-Bayesian joint estimation of systematic noise and the transit signal as a mitigating strategy for analogous overfitting, in their case primarily to address systematic errors due to pointing errors in the K2 mission. We echo their conclusion that these approaches have clear advantages in transiting exoplanet detection. 

The Bayesian joint detectors described in the current work are computationally expensive (Section \ref{sec:complexity}) however we have demonstrated that such detectors can be used effectively in conjunction with the phase correlation method \citep{phase} applied to cotrended light curves. Phase correlation reduces the size of the transit signal parameter search space significantly and allows freed-up computational resources to be allocated to finer searches over other transit parameters such as orbital period. We note that it is possible that the use of phase correlation on cotrended light curves may have introduced a slight bias in favor of the standard method detector as the phase correlation method finds the optimal phase estimate for a cotrended lightcurve. However, we do not believe this affects our conclusions from the current work. We note also that in future, we propose to explore the use of ranked cross-correlation between the light curves and candidate transit signals in a generalized approach to optimize the identification of transit epochs.

As described in Section \ref{sec:design_impl} the joint detection framework presented here admits many implementation choices and optimizations. As an exploratory evaluation of the statistical performance of these Bayesian joint detectors we took care to maintain consistency between the implementations to allow meaningful relative comparisons but did not fine-tune the detectors to produce optimal detection rates. For example the same priors and epoch estimates were used across all models. Also the detection efficiency is computed over a single observing quarter of Kepler data; in practice a folded test statistic across multiple quarters would likely be less vulnerable to poor data quality in a single quarter. Similarly, we have not yet evaluated adjunct techniques to enhance detection efficiency, including methods such as outlier detection or harmonic filtering. These alternative implementation choices will be explored in future work.

The initial joint detector feasibility test with Kepler data containing known exoplanet detections is described in Section \ref{sec:kepler_results} and the results depicted in Figure \ref{fig:histp}. These preliminary tests show that detectors A and B were able to recover known exoplanets at a rate comparable or marginally superior to the standard model. All detectors show a large number of spurious detections, particularly at larger orbital periods within the search window. The spurious detections are primarily due to the short data segment used and the small sample of lightcurves from which the systematic noise prior is built. However, we also believe that the spurious long-period detections may be reduced with more careful optimization of the detectors for maximal detection efficiency. This was not within the scope of the current work. Specifically we believe that the lack of outlier rejection may be a contributing factor to the spurious long-period detections. These detections may also be reduced simply by utilizing more data over longer observational periods. We stress however that the joint detection tests with Kepler data containing exoplanets is preliminary in nature and primarily, though successful, an initial feasibility test. 

\section{Conclusions}\label{sec:conclusion}
We have developed a Bayesian framework for the joint detection of systematic noise and exoplanet transit signals. We formulated our detection framework as a likelihood ratio test and used a Neyman-Pearson optimality criterion. Two general Bayesian approaches were used, namely maginalization over the systematic noise (detector A) and conditional estimation of the systematic noise (detector B). Under the assumption of a Gaussian prior for the systematic noise we show that these detectors can be expressed in closed form as matched filters. The performance of the joint detectors was evaluated in numerical recovery tests of injected transit signals added to raw Kepler light curves. The Kepler data in the injection tests excluded known exoplanet detections. Further, an initial feasibility test was performed by applying the detectors to a subset of Kepler data from which confirmed exoplanet detections had not been excluded. An additional standard detector which performed sequential cotrending and detection was defined as a comparator during the numerical tests.

The principal conclusions of the paper are as follows:
\begin{itemize}

\item{In the injection tests the joint Bayesian detectors A and B show an improvement of $\sim 2\%$ in overall detection efficiency relative to the standard detector. As an initial exploratory assessment, without significant detector optimization, the joint detectors therefore show sufficient promise to warrant further detailed investigation. We have identified several proposed detector efficiency optimizations.}

\item{The joint detectors A and B show specific improvement in detection efficiency for exoplanets with both short orbital periods ($P < 10 {\rm d})$ and low ratios of planet to host star radius $\frac{R_p}{R_*} (< 0.05\%)$. We conclude that joint estimation offers improved separation of residual high-frequency systematic noise and overlapping transit signals and mitigates overfitting. We believe this approach has future potential in this regime specifically as well as low S/N environments.}

\item{The Bayesian joint detectors are computationally expensive but we have shown that they are tractable with contemporary high-performance computing resources. The Bayesian approach offers the advantage of full statistical generality regarding the form of the probability distribution adopted for the statistical noise and stellar signal and the parameter estimators used. We have demonstrated that phase correlation can be used in conjunction with this method to reduce significantly the transit parameter search space and thereby the net computational complexity.}

\end{itemize}

\section{Acknowledgements} 
This research is part of the Blue Waters sustained-petascale computing project, which is supported by the National Science Foundation (awards OCI-0725070 and ACI-1238993) and the state of Illinois. Blue Waters is a joint effort of the University of Illinois at Urbana-Champaign and its National Center for Supercomputing Applications. This paper includes data collected by the Kepler mission. Funding for the Kepler mission is provided by the NASA Science Mission directorate.

\appendix
\section{Phase Correlation Method}\label{sec:phasecorr} 
The phase correlation method \citep{phase} is a classic image registration technique. In the current work this method is used to estimate the epoch $t_0$ of a candidate transit signal from a coarsely cotrended lightcurve. This method estimates the offset $t_0$ between a signal $\bold{x}(t)$ and a shifted version of this same signal $\bold{x}(t - t_0)$. Denote the Fourier transform of $\bold{x}(t)$ as $\bold{X}(\omega)=\mathcal{F}\{\bold{x}(t)\}$. The Fourier shift theorem \citep{bracewell2000fourier} yields:
\begin{align}
    \mathcal{F}\{\bold{x}(t - t_0)\} = \bold{X}(\omega)e^{-j \omega t_0}
\end{align}
The pixel-level phase correlation method \citep{phase} uses this property to form the normalized cross-power spectrum as an estimate of this phase shift: 
\begin{align}
    \frac{\bold{X}(\omega)\bold{X}^*(\omega)e^{j \omega t_0}}{|\bold{X}(\omega)||\bold{X}(\omega)e^{-j \omega t_0}|}= e^{j \omega t_0}
\end{align}
The inverse Fourier transform of this expression yields a delta function centered on the position of the shift $t_0$, thus providing a local maximum. This method is also effective if the observed signal contains additive noise or is improperly scaled \citep{phase}.

\section{Extension to Multiple Quarters}\label{sec:multi-quarter}
In this Appendix we discussion the extension of our current detectors to the case of multiple observing quarters, although we stress that these methods were not applied in the current work.

The complexity of the detectors grows polynomially with $N$ as described in Section ~\ref{sec:complexity}. As such, application to long time series data becomes prohibitively expensive. We will briefly outline how the detectors may be applied to multi-quarter searches with linear growth in complexity.

The form of the generalized matched filter described in Section \ref{sec:matched_filter} depends strongly on the form of the signal covariance matrix. If we assume data between quarters to be uncorrelated, a covariance matrix describing all observations will be of block-diagonal form. For $Q$ quarters, where each quarter has covariance matrix $\Cov_q: q \in Q$, we obtain a multi-quarter covariance matrix $\Cov_z$:
\begin{equation}
   \Cov_z =  \begin{bmatrix}
    \Cov_1 \\
    & \Cov_2 \\
    && \ddots \\
    &&& \Cov_Q
    \end{bmatrix}
\end{equation}

This allows a linear decomposition of the matched filter:
\begin{align}
    \frac{\bold{y}^T \Cov_n^{-1} \bold{t}}{\sqrt{\bold{t}^T \Cov_n^{-1} \bold{t}}} = \frac{\sum_{q \in Q} \bold{y}_q^T \Cov_q^{-1} \bold{t}_q}{\sqrt{\sum_{q \in Q} \bold{t}_q^T \Cov_q^{-1} \bold{t}_q}}
\end{align}

In this form, the total computational cost is the complexity of a single quarter multiplied by the total number of quarters.

Given the definition of detector A in Section ~\ref{sec:deta} under the systematic noise model of Section ~\ref{sec:sysnoise}, and the stellar models in Section ~\ref{sec:stellar}, the multi-quarter covariance matrix $\Cov_z$ will in fact be in block diagonal form. For detector B, further optimization is possible due to the the Toeplitz structure of stellar covariance within a given quarter. In addition, for detector B, the Toeplitz nature of $\Cov_q$, allows each term within the summand to be computed efficiently in the Fourier domain.
\bibliography{mybib}{}
\bibliographystyle{aasjournal}


\end{document}